\documentclass{emulateapj}

\usepackage{natbib}
\usepackage{amsmath}

\newcommand{\ugriz}{\protect\hbox{$ugriz$} }
\newcommand{\ugr}{\protect\hbox{$ugr$} }
\newcommand{\gri}{\protect\hbox{$gri$} }

\newcommand{\ubv}{\protect\hbox{$U\!BV$} }
\newcommand{\about}{$\sim\!\!$~}

\newcommand{\be}{\begin{displaymath}}
\newcommand{\ee}{\end{displaymath}}
\newcommand{\err}[2]{\ensuremath{^{+#1}_{-#2}}}
\def\lsim{\hbox{\rlap{\raise 0.425ex\hbox{$<$}}\lower 0.65ex\hbox{$\sim$}}}
\def\gsim{\hbox{\rlap{\raise 0.425ex\hbox{$>$}}\lower 0.65ex\hbox{$\sim$}}}

\def\arcdeg{\mbox{$^\circ$}}

\newcommand{\kms}{km~s$^{-1}$ }

\newcommand{\dof}{\rm dof}
\newcommand{\mean}[1]{\left \langle #1 \right \rangle}

\shorttitle{UV Mismatch in SNe~Ia}
\shortauthors{Foley et~al.}

\begin{document}

\title{A Mismatch in the Ultraviolet Spectra between Low-Redshift and
Intermediate-Redshift Type Ia Supernovae as a Possible Systematic
Uncertainty for Supernova Cosmology}

\def\cfa{1}
\def\berk{2}
\def\clay{3}
\def\chikav{4}
\def\chiast{5}
\def\cape{6}
\def\saao{7}
\def\aims{8}
\def\fermi{9}
\def\nd{10}
\def\rut{11}
\def\tok{12}
\def\port{13}
\def\jhu{14}
\def\stsci{15}
\def\penn{16}
\def\psu{17}
\def\psug{18}
\def\stock{19}

\author{
{Ryan~J.~Foley}\altaffilmark{\cfa,\berk,\clay},
{Alexei~V.~Filippenko}\altaffilmark{\berk},
{Richard~Kessler}\altaffilmark{\chikav,\chiast},
{Bruce~Bassett}\altaffilmark{\cape,\saao,\aims},
{Joshua~A.~Frieman}\altaffilmark{\chikav,\chiast,\fermi},
{Peter~M.~Garnavich}\altaffilmark{\nd},
{Saurabh~W.~Jha}\altaffilmark{\rut},
{Kohki~Konishi}\altaffilmark{\tok},
{Hubert~Lampeitl}\altaffilmark{\port},
{Adam~G.~Riess}\altaffilmark{\jhu,\stsci},\\
{Masao~Sako}\altaffilmark{\penn},
{Donald~P.~Schneider}\altaffilmark{\psu,\psug},
{Jesper~Sollerman}\altaffilmark{\stock}, and
{Mathew~Smith}\altaffilmark{\cape}
}

\altaffiltext{\cfa}{
Harvard-Smithsonian Center for Astrophysics,
60 Garden Street, 
Cambridge, MA 02138, USA.
}
\altaffiltext{\berk}{
Department of Astronomy,
University of California,
Berkeley, CA 94720-3411, USA.
}
\altaffiltext{\clay}{
Clay Fellow. Electronic address rfoley@cfa.harvard.edu .
}
\altaffiltext{\chikav}{
Kavli Institute for Cosmological Physics, 
The University of Chicago,
5640 South Ellis Avenue,
Chicago, IL 60637, USA.
}
\altaffiltext{\chiast}{
Department of Astronomy and Astrophysics,
The University of Chicago,
5640 South Ellis Avenue,
Chicago, IL 60637.
}
\altaffiltext{\cape}{
Department of Mathematics and Applied Mathematics,
University of Cape Town,
Rondebosch 7701, South Africa.
}
\altaffiltext{\saao}{
South African Astronomical Observatory,
P.O. Box 9,
Observatory 7935, South Africa.
}
\altaffiltext{\aims}{
African Institute for Mathematical Sciences,
6--8 Melrose Road,
Muizenberg, Cape Town, South Africa.
}
\altaffiltext{\fermi}{
Center for Particle Astrophysics, 
Fermi National Accelerator Laboratory,
P.O. Box 500,
Batavia, IL 60510, USA.
}
\altaffiltext{\nd}{
University of Notre Dame,
225 Nieuwland Science,
Notre Dame, IN 46556-5670, USA.
}
\altaffiltext{\rut}{
Department of Physics and Astronomy,
Rutgers University,
136 Frelinghuysen Road,
Piscataway, NJ 08854, USA.
}
\altaffiltext{\tok}{
Institute for Cosmic Ray Research,
University of Tokyo,
5-1-5, Kashiwanoha,
Kashiwa, Chiba, 277-8582, Japan.
}
\altaffiltext{\port}{
Institute of Cosmology and Gravitation,
Mercantile House,
Hampshire Terrace,
University of Portsmouth,
Portsmouth PO1 2EG, UK.
}
\altaffiltext{\jhu}{
Department of Physics and Astronomy,
Johns Hopkins University,
Baltimore, MD 21218, USA.
}
\altaffiltext{\stsci}{
Space Telescope Science Institute,
3700 San Martin Drive,
Baltimore, MD 21218, USA.
}
\altaffiltext{\penn}{
Department of Physics and Astronomy,
University of Pennsylvania,
209 South 33rd Street,
Philadelphia, PA 19104, USA.
}
\altaffiltext{\psu}{
Department of Astronomy and Astrophysics,
525 Davey Laboratory,
Pennsylvania State University,
University Park, PA 16802, USA.
}
\altaffiltext{\psug}{
Institute for Gravitation and the Cosmos,
The Pennsylvania State University,
University Park, PA 16802, USA.
}
\altaffiltext{\stock}{
Oskar Klein Centre,
Department of Astronomy,
Stockholm University,
106 91 Stockholm, Sweden.
}

\begin{abstract}
We present Keck high-quality rest-frame ultraviolet (UV) through
optical spectra of 21 Type Ia supernovae (SNe~Ia) in the redshift
range $0.11 \le z \le 0.37$ and a mean redshift of 0.22 that were
discovered during the Sloan Digital Sky Survey-II (SDSS-II) SN Survey.
Using the broad-band photometry of the SDSS survey, we are able to
reconstruct the SN host-galaxy spectral energy distributions (SEDs),
allowing for a correction for the host-galaxy contamination in the
SN~Ia spectra.  Comparison of composite spectra constructed from a
subsample of 17 high-quality spectra to those created from a
low-redshift sample with otherwise similar properties shows that the
Keck/SDSS SNe~Ia have, on average, extremely similar rest-frame
optical spectra but show a UV flux excess.  This observation is
confirmed by comparing synthesized broad-band colors of the individual
spectra, showing a difference in mean colors at the 2.4--4.4$\sigma$
level for various UV colors.  We further see a slight difference in
the UV spectral shape between SNe with low-mass and high-mass host
galaxies.  Additionally, we detect a relationship between the flux
ratio at 2770 and 2900~\AA\ and peak luminosity that differs from that
observed at low redshift.  We find that changing the UV SED of an
SN~Ia within the observed dispersion can change the inferred distance
moduli by \about 0.1~mag.  This effect only occurs when the data probe
the rest-frame UV.  We suggest that this discrepancy could be due to
differences in the host-galaxy population of the two SN samples or to
small-sample statistics.
\end{abstract}

\keywords{supernovae---general, cosmology---observations, distance scale}

\defcitealias{Foley08:uv}{FFJ}
\defcitealias{Kessler09:cosmo}{K09}


\section{Introduction}\label{s:intro}

Type Ia supernovae (SNe~Ia), being both extremely luminous and having
well-calibrated peak luminosities \citep[starting with the work
of][]{Phillips93}, are excellent and precise distance indicators at
cosmological scales; see, for example, \citet{Leibundgut01},
\citet{Livio01}, \citet{Perlmutter03}, and \citet{Filippenko05a} for
reviews.  Measurements of several hundred SNe~Ia at low redshifts
\citep[e.g., ][]{Hamuy96:lc, Riess99:lc, Jha06:lc, Hicken09:lc,
Contreras10, Ganeshalingam10, Stritzinger11} and high redshifts have
shown that the expansion rate of the Universe is currently
accelerating \citep{Riess98:Lambda, Perlmutter99}, and that the
equation-of-state parameter of the dark energy, $w = P/(\rho c^{2})$,
is consistent with the cosmological constant, $w = -1$
\citep[e.g., ][]{Riess07, Wood-Vasey07, Hicken09:de, Kessler09:cosmo,
Conley11, Sullivan11}.  The underlying assumption of this work is that
we can accurately measure distances to SNe~Ia at high redshifts using
relationships between distances and observations of SNe~Ia at low
redshifts.

If there are physical differences between low-redshift and
high-redshift SNe~Ia (in a way that affects our derived luminosity
distances), we would expect to notice discrepancies in their observed
spectra.  Theoretical studies have determined that changing progenitor
metallicity, a likely difference between low-redshift and
high-redshift SNe~Ia, produces relatively small variations in the
optical spectra of SNe~Ia but larger differences in their ultraviolet
(UV) spectra \citep[e.g.,][]{Hoflich98, Lentz00, Sauer08}.
\citet{Foley08:comp} and \citet{Ellis08} have recently shown that
high-redshift SNe~Ia are similar to their low-redshift counterparts,
though both studies had major limitations: an inability to properly
correct for host-galaxy contamination in the high-redshift SN spectra
\citep{Foley08:comp} and an insufficient low-redshift comparison
sample \citep{Ellis08}.

\citet{Sullivan09} compared composite spectra from three samples of
SNe~Ia with mean redshifts ($z$) of 0.02, 0.48, and 1.16.  The spectra
in the samples had (by construction) similar phases, but the SNe had
slightly different stretches (consistent with the trend with redshift;
\citealt{Howell07}) and colors.  The composite spectra differed such
that the strength of features corresponding to intermediate-mass
elements (IMEs) are weaker at higher redshifts.  \citet{Sullivan09}
argue that the changing stretch distribution with redshift results in
SN~Ia explosions producing more $^{56}$Ni, and thus having less mass
in IMEs, at higher redshift.  \citet{Sullivan09} also found that their
low-redshift composite spectrum had less flux for $\lambda <
3500$~\AA, but they dismiss the result since their low-redshift sample
contains only three SNe~Ia with spectra covering these wavelengths.
\citet{Cooke11} created a composite spectrum from 10 nearby
low-redshift SNe~Ia and also found that the low-redshift SNe had less
flux than the high-redshift SNe in the UV.  Both the low and
high-redshift samples had similar light-curve shape distributions,
making it unlikely that the discrepancy is caused by differences in
the population.  Nonetheless, they decided not to examine this feature
until they obtained more data.

Further indications of a possible change in the UV properties of
SNe~Ia come from the analysis of \citet[hereafter
K09]{Kessler09:cosmo}, who found differences in the rest-frame
$U$-band data for low-redshift and high-redshift samples.  Using only
the Sloan Digital Sky Survey-II (SDSS-II) sample to extract
cosmological parameters, exclusion of the rest-frame UV region caused
a systematic offset in the measurement of $w$ by \about 0.3.
(However, this offset is significantly reduced when combining with
higher-redshift samples.)  This difference is referred to as the
``$U$-band anomaly'' by \citetalias{Kessler09:cosmo}.  The differences
could be caused by calibrations (the observer-frame $U$ band is
notoriously difficult to properly calibrate), or by real differences
in the samples.  Since the inclusion of the $U$-band and SDSS $u$-band
data had a very different effect on the low-redshift sample and the
lowest-redshift SDSS-II SNe ($z < 0.1$), respectively,
\citetalias{Kessler09:cosmo} suggested that the low-redshift,
non-SDSS-II sample was the cause of the effect, which could easily be
the result of incorrect calibration.

\citet*[][hereafter FFJ]{Foley08:uv}  showed that for a sample of six
low-redshift SNe~Ia, the steepness of their UV spectra near maximum
brightness, the ``UV ratio'' [$\mathcal{R}_{UV} = f_{\lambda}
(2770~\text{\AA}) / f_{\lambda} (2900~\text{\AA})$], appears to be
highly correlated with peak luminosity.  Although this relationship
could provide new insights into the UV properties of SNe~Ia and reduce
the scatter in their calibrated absolute luminosities, it is based on
few SNe (but additional low-redshift SNe~Ia are consistent with the
relationship; \citealt{Bufano09}; \citealt{Foley12:09ig}).  Moreover,
the relationship has not yet been tested at high redshifts.

The SDSS-II SN survey has discovered approximately 1000 SNe~Ia at $z
\lesssim 0.4$ with \about 500 being spectroscopically confirmed
\citep{Frieman08, Sako08, Kessler09:cosmo, Sollerman09,
Lampeitl10:cosmo, Sako11}.  We describe this redshift range as
``intermediate,'' but note that if one must choose a binary
description, it would naturally fall in the ``high-redshift'' category
rather than the ``low-redshift'' one.  As a complementary program at
the Keck Observatory to examine potential differences in the UV
spectral-energy distributions (SEDs) of low-redshift and high-redshift
SNe~Ia and to test the validity of the \citetalias{Foley08:uv}
relationship at high redshift, we observed 21 SDSS-II SNe~Ia (the
``Keck/SDSS'' sample) with the intention of obtaining high
signal-to-noise ratio (S/N) rest-frame UV through optical spectra.  We
have used the deep multi-band photometry of the SN host galaxies from
SDSS to correct for host-galaxy contamination.  Utilizing the large,
high quality, low-redshift SN~Ia samples from \citet{Foley08:comp}
(described in greater detail by \citealt{Foley08:comp} and
\citetalias{Foley08:uv}), we have been able to perform a better
comparison of low and high-redshift SN~Ia spectra.  Note that the
low-redshift composite spectra presented by \citet{Foley08:comp} were
created using the same methods as those employed in this paper.
\citet{Hsiao07} also constructed a template spectrum similar to our
low-redshift composite spectra \citep{Foley08:comp}; however, the
\citet{Hsiao07} template spectra contain high-redshift SN~Ia data and
hence are not ideal for evolutionary studies.

This paper is structured in the following way.  Section~\ref{s:obs}
discusses our sample of SNe~Ia, the Keck spectroscopic observations,
and the data reduction.  We also describe how we correct for
host-galaxy contamination and our method for creating composite
spectra.  In Section~\ref{s:comp}, we compare our Keck/SDSS sample to
the low-redshift sample presented by \citet{Foley08:comp}.
Section~\ref{s:cosmo} explores the implications of our findings for SN
cosmology.  In Section~\ref{s:subsamples}, we split the Keck/SDSS
sample by various observational properties to determine their effect
on spectra.  We examine the UV ratio in Section~\ref{s:uvratio}.
Section~\ref{s:disc} explores potential systematic errors.  We
summarize our conclusions in Section~\ref{s:conc}, including some
discussion for future research.


\section{Observations and Data Reduction}\label{s:obs}

The SDSS-II survey was a three-year extension of the SDSS (for
technical descriptions of the SDSS and its data products, see
\citealt{York00} and \citealt{Stoughton02}).  The SDSS-II SN Survey,
one component of the SDSS-II program, was carried out over three
seasons, each of duration three months (Sep.--Nov.\ 2005--2007).
Using the SDSS wide-field imaging camera \citep{Gunn98} on the SDSS
2.5~m telescope \citep{Gunn06}, it was a time-domain survey in \ugriz
over 300 square degrees centered on the celestial equator
\citep{Fukugita96}.  The SN survey took advantage of the excellent
photometric and astrometric calibration achieved by the SDSS
\citep{Hogg01, Smith02:SDSSfilters, Pier03, Ivezic04, Tucker06}.  Its
main purpose was to fill in the ``redshift desert" ($ 0.1 < z < 0.3$)
in the existing SN~Ia Hubble diagram \citep[which was visible in SN~Ia
Hubble diagrams at the time of the creation of the project;
e.g.,][]{Riess04}, with a focus on reducing systematic errors in SN~Ia
distance estimates.  In total, the SDSS-II SN Survey discovered
\about 500 spectroscopically confirmed SNe~Ia out to a redshift of
\about 0.4 \citep{Frieman08}.  The first cosmological results were
presented by \citetalias{Kessler09:cosmo}, with additional
investigations by \citet{Sollerman09} and \citet{Lampeitl10:cosmo}.

SNe for the Keck/SDSS sample were detected as part of the SDSS-II SN
survey \citep{Barentine05, Bassett06, Bassett07a, Bassett07b,
Bassett07c}.  In addition to many spectroscopic observations of SNe
with smaller telescopes \citep{Zheng08, Ostman11, Konishi11:subaru}, a
specific program was developed at the Keck Observatory to observe the
rest-frame UV spectra of $z \approx 0.3$ SNe~Ia.  Targets were
selected in a manner described by \citet{Sako08}; objects which
appeared to have the appropriate colors and light curves of an SN~Ia
near maximum light at $z \approx 0.3$ were chosen for observation in
our program.  Some host galaxies had predetermined redshifts, which
would occasionally influence the decision to observe an object.  About
half of the targets were already spectroscopically confirmed as SNe~Ia
near maximum brightness in the appropriate redshift range by other
telescopes before observations were made at Keck.  No additional data
(e.g., host-galaxy properties, brightness, etc.) were used in deciding
which objects were observed.

\subsection{Photometry}

Precise SDSS-II SN photometry was obtained using the ``Scene Modeling
Photometry'' (SMP) technique, developed for this purpose and described
in detail by \citet{Holtzman08}.  The basic idea of SMP is to model
the data as a time-varying point source (the SN) and sky background
along with a time-independent galaxy background, all convolved with a
time-varying point-spread function (PSF).  The model is constrained by
a global fit that uses all of the images taken during the survey that
cover the SN region.  Each image is modeled as a sum of fluxes from
the SN, host galaxy, sky background, and calibration stars near the
SN.  The calibration stars are taken from the SDSS catalog produced by
\citet{Ivezic07}.  The fitted parameters are SN position, SN flux for
each epoch and passband, and the host-galaxy intensity distribution in
each passband.  The galaxy model for each passband is a $20 \times 20$
grid (with a grid scale set by the CCD pixel scale, $0.\farcs4 \times
0.\farcs4$) in sky coordinates, and each of the $400 \times 5 = 2000$
intensities is an independent fit parameter.  In addition to the fit
parameters, the flux model also depends on measured quantities that
include sky background, magnitudes of calibration stars,
position-dependent PSF, and a scaling between calibrated flux and CCD
counts.  There is no pixel resampling or image convolution, resulting
in correct statistical uncertainties.

\subsection{Spectroscopy}\label{ss:spec}

Spectra of SDSS-II targets were obtained with the Keck-I 10~m
telescope using the low-resolution imaging spectrometer \citep[LRIS;
][]{Oke95}.  Observations in 2005 were obtained with the 600/4000
grism and the 400/8500 grating with the polarimetry optics in place.
Observations in 2006 and 2007 were obtained with the 400/3400 grism
without the polarimetry optics, resulting in better response at blue
wavelengths.  Additionally, an atmospheric dispersion compensator
(ADC) was in place for observations in 2007.  This resulted in spectra
covering the entire optical range, approximately 3400--9400~\AA.  All
spectra were obtained at relatively low airmass, and several standard
stars were observed at similar airmass throughout the night.  To
reduce differential slit losses due to atmospheric dispersion
\citep{Filippenko82}, the spectra were obtained with a position angle
near parallactic, except for some objects in 2007 which were observed
at low airmass and with the ADC.

Additional spectra of a subset of the host galaxies were obtained with
the MagE spectrograph \citep{Marshall08} on the Magellan Clay 6.5~m
telescope in December 2008, well after the SNe had faded.  The slit
was placed at the position of the SN with a position angle that went
through the host nucleus.  Because of this alignment, we were able to
extract spectra at both positions.

Standard CCD processing and spectrum extraction of the LRIS data were
accomplished with IRAF\footnote{IRAF: the Image Reduction and Analysis
Facility is distributed by the National Optical Astronomy Observatory,
which is operated by the Association of Universities for Research in
Astronomy, Inc. (AURA) under cooperative agreement with the National
Science Foundation (NSF).}.  The data were extracted using the optimal
algorithm of \citet{Horne86}.  Low-order polynomial fits to
calibration-lamp spectra were used to establish the wavelength scale.
Small adjustments derived from night-sky lines in the object frames
were applied.  For the MagE spectra, the sky was subtracted from the
images using the method described by \citet{Kelson03}.  We employed
IRAF and our own routines to flux calibrate the data and remove
telluric lines using the well-exposed continua of the
spectrophotometric standards \citep{Wade88, Foley03, Foley09:08ha}.
We used the spectrophotometric standards BD+284211 and Feige~34
\citep{Oke90} to calibrate the spectra from the blue arm and BD+174708
and HD~19445 \citep{Oke83} to calibrate the red arm.  Our relative
photometric flux calibration is typically accurate to 5\% across our
entire wavelength range (\citealt{Matheson08}; Silverman et~al.,
submitted).

\begin{figure}
\begin{center}
\epsscale{1.75}
\rotatebox{90}{
\plottwo{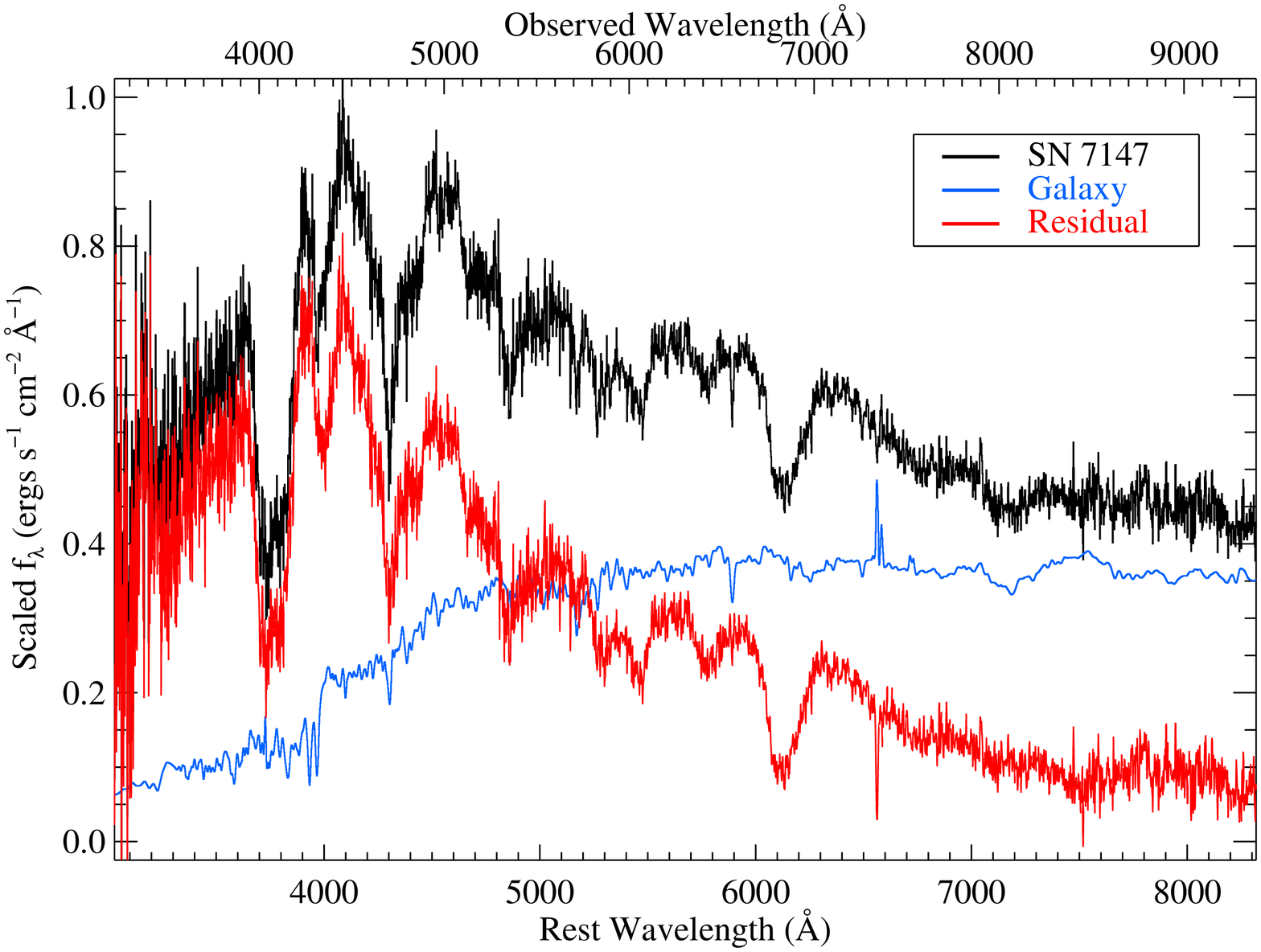}{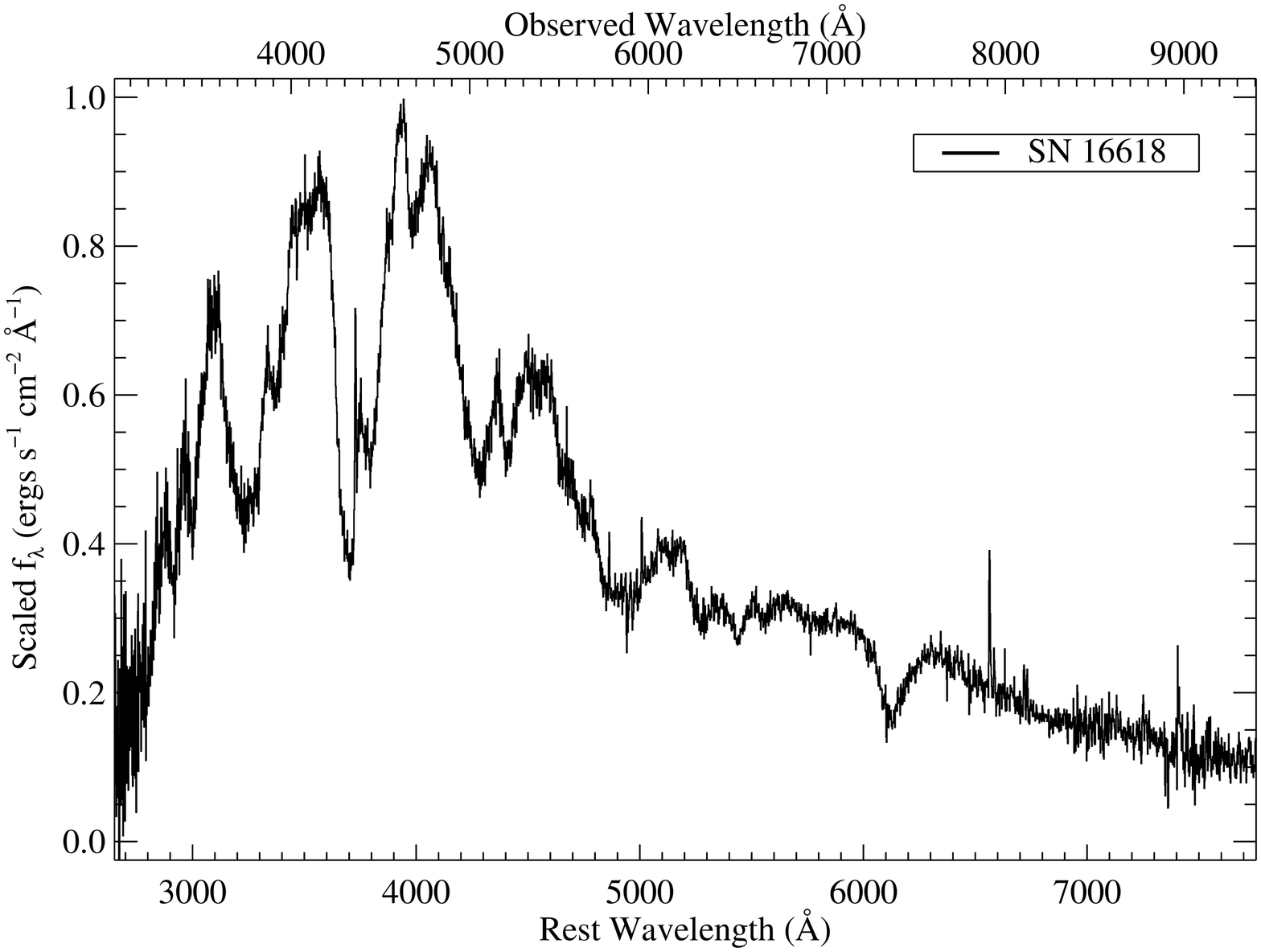}}
\caption{Spectra of SNe~16618 (top) and 7147 (bottom), examples of
SNe with no and some host-galaxy contamination, respectively.  The
black curves are the observed spectra.  The blue curve for SN~7147 is
the reconstructed galaxy spectrum of its host.  The red curve is the
residual spectrum after galaxy-contamination correction.  The
galaxy-subtraction method produces clean SN spectra.  See the
electronic version for figures of all spectra of SNe~Ia used in this
study.}\label{f:spec}
\end{center}
\end{figure}

\begin{figure}
\begin{center}
\epsscale{1.95}
\rotatebox{90}{
\plotone{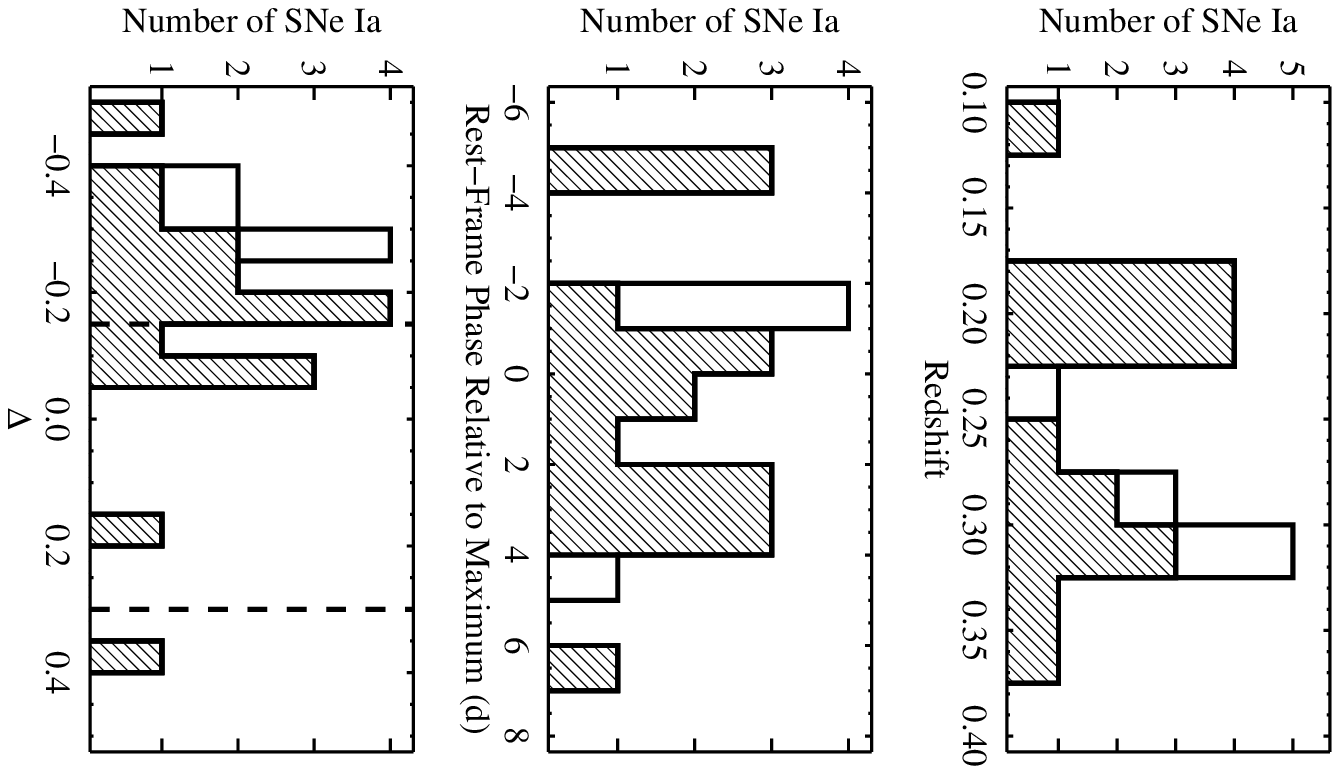}}
\caption{Histogram of the redshift (top panel), rest-frame phase as
determined from the light curves (middle panel), and $\Delta$
distributions (bottom panel) for the Keck/SDSS SNe~Ia presented in
this paper.  The dashed and open histograms indicate the SNe in the
Nominal sample and additional SNe in the All sample, respectively; see
text (Section~\ref{ss:composite}) for a description of these two
samples.  The dashed lines in the bottom panel mark the regions of
high-luminosity ($\Delta < -0.15$), normal ($-0.15 < \Delta < 0.3$),
and low-luminosity ($\Delta > 0.3$) SNe, as defined by
\citet{Jha07}.}\label{f:hist}
\end{center}
\end{figure}

The Keck/SDSS sample presented in this paper consists of 21 SNe~Ia.
The properties of the SNe and their spectra are listed in
Table~\ref{t:spec}.  Two individual spectra are presented in
Figure~\ref{f:spec} (see the online version for all individual
spectra).  Histograms of the sample's redshift, rest-frame phase $t$,
and $\Delta$\footnote{$\Delta$ is a unitless parameter that describes
light-curve shape and is related to luminosity, with smaller values of
$\Delta$ corresponding higher luminosity; $M_{V} (t = 0) = -19.504 +
0.736 \Delta + 0.182 \Delta^{2} + 5 \log (H_{0}/65 {\rm
~km~s}^{-1}{\rm ~Mpc}^{-1})$~mag; $\Delta = 0$ corresponds to $\Delta
m_{15} (B) = 1.07$~mag and stretch $s_{B} = 0.96$ \citep{Riess96,
Jha07}.} are presented in Figure~\ref{f:hist}.  It is worth noting
that the low and high-redshift light curves were fit with different
versions of the multicolor light-curve shape method
\citep[MLCS;][]{Riess96}: at low redshift by the publicly available
version detailed by \citet{Jha07}, and at high redshift by a slightly
modified internal SDSS-II version. The differences between the
implementations are minor, however, and the derived values are
essentially the same.

\begin{center}
\begin{deluxetable*}{l@{ }l@{ }c@{ }l@{ }c@{ }c@{ }c@{ }c@{ }c@{ }c@{ }c}
\tabletypesize{\scriptsize}
\tablewidth{0pt}
\tablecaption{Supernova Information\label{t:spec}}
\tablehead{
\colhead{IAU SN} &
\colhead{SDSS-II} &
\colhead{UT Date of Spectrum} &
\colhead{} &
\colhead{} &
\colhead{Galaxy} &
\colhead{Rest-Frame} &
\colhead{} &
\colhead{Host $A_{V}$} &
\colhead{Host Mass} &
\colhead{Host} \\
\colhead{Name} &
\colhead{ID} &
\colhead{yyyy-mm-dd} &
\colhead{Redshift} &
\colhead{Airmass} &
\colhead{Cont. (\%)\tablenotemark{a}} &
\colhead{Phase (days)\tablenotemark{b}} &
\colhead{$\Delta$\tablenotemark{c}} &
\colhead{(mag)\tablenotemark{d}} &
\colhead{($10^{9}\, {\rm M}_{\sun}$)\tablenotemark{e}} &
\colhead{$B1000$\tablenotemark{f}}}

\startdata

2005jk & 6304  & 2005-11-05.521 & 0.19   & 1.55 &     0.0 & \phs6.3 (0.4) &  $-0.06$ (0.11) & 0.41 (0.06) &   20.   &    0.33 \\
2005ix & 6315  & 2005-11-05.253 & 0.26   & 1.19 &    30.1 & \phs2.9 (0.6) &  $-0.15$ (0.08) & 0.01 (0.02) &    1.7  &     3.2 \\
2005jd & 6649  & 2005-11-05.563 & 0.32   & 1.83 &     1.6 & \phs2.7 (0.5) &  $-0.26$ (0.09) & 0.02 (0.02) &    4.4  &    52   \\
2005ik & 6699  & 2005-11-05.292 & 0.32   & 1.20 &     0.0 & \phs3.0 (0.6) &  $-0.26$ (0.13) & 0.06 (0.20) &    3.6  &    55   \\
2005jc & 6933  & 2005-11-05.478 & 0.215  & 1.54 &     4.5 & \phs0.8 (0.3) &  $-0.17$ (0.09) & 0.22 (0.07) & \nodata & \nodata \\
2005jl & 6936  & 2005-11-05.269 & 0.180  & 1.14 &     7.4 &  $-0.1$ (0.3) &  $-0.20$ (0.08) & 0.20 (0.06) &    7.0  &     4.1 \\
2005jh & 7147  & 2005-11-05.418 & 0.111  & 1.60 &    32.3 & \phs0.0 (0.2) & \phs0.36 (0.07) & 0.01 (0.02) &   12.   &     5.6 \\
2005jm & 7243  & 2005-11-05.319 & 0.204  & 1.25 &    13.3 &  $-4.6$ (0.1) &  $-0.22$ (0.11) & 0.17 (0.08) &    0.31 &    46   \\
2005ji & 7473  & 2005-11-05.436 & 0.214  & 1.37 &     5.6 & \phs1.0 (0.4) &  $-0.07$ (0.09) & 0.02 (0.03) &    2.0  &    67   \\
2005jn & 7475  & 2005-11-05.457 & 0.33   & 1.51 &     0.0 & \phs3.3 (0.5) &  $-0.16$ (0.08) & 0.01 (0.01) & \nodata & \nodata \\
2005jp & 7847  & 2005-11-05.544 & 0.2128 & 1.64 &    32.1 &  $-1.3$ (0.3) &  $-0.11$ (0.13) & 0.43 (0.08) &   13.   &     3.6 \\
2006pf & 16567 & 2006-11-23.470 & 0.3656 & 1.55 &    40.4 & \phs2.0 (0.8) &  $-0.32$ (0.10) & 0.01 (0.02) &   12.   &    17   \\
2006pq & 16618 & 2006-11-23.447 & 0.193  & 1.48 &     0.0 &  $-4.4$ (0.5) &  $-0.36$ (0.13) & 0.08 (0.07) &    4.0  &    51   \\
2007my & 19027 & 2007-10-16.244 & 0.293  & 1.10 &     0.0 & \phs3.1 (0.8) & \phs0.18 (0.16) & 0.04 (0.20) &    0.77 &    47   \\
2007lu & 19029 & 2007-10-14.244 & 0.319  & 1.13 &    25.5 &  $-0.1$ (1.0) &  $-0.47$ (0.07) & 0.02 (0.03) &    2.6  &    63   \\
2007ml & 19101 & 2007-10-16.263 & 0.19   & 1.57 &     0.0 &  $-4.0$ (0.3) &  $-0.08$ (0.09) & 0.32 (0.07) & \nodata & \nodata \\
2007lw & 19128 & 2007-10-14.267 & 0.29   & 1.30 &     0.0 & \phs1.5 (0.7) &  $-0.23$ (0.09) & 0.02 (0.04) & \nodata & \nodata \\
\hline
2005jo & 7512  & 2005-11-05.615 & 0.23   & 1.89 & \nodata &  $-1.0$ (0.4) &  $-0.27$ (0.09) & 0.26 (0.07) & \nodata & \nodata \\
2006pt & 16644 & 2006-11-23.495 & 0.2990 & 1.49 & \nodata &  $-1.9$ (1.3) &  $-0.34$ (0.17) & 0.16 (0.20) &    36.  &     8.6 \\
2006pz & 16789 & 2006-11-23.525 & 0.3250 & 1.54 & \nodata & \phs5.6 (1.1) &  $-0.36$ (0.17) & 0.15 (0.15) &   180   &     1.1 \\
2007qu & 20829 & 2007-11-12.298 & 0.31   & 1.14 & \nodata &  $-1.2$ (0.4) &  $-0.25$ (0.14) & 0.01 (0.02) &    30.  &     2.1

\enddata

\tablecomments{SNe below the line are those rejected from the
Nominal sample.  See Section~\ref{ss:composite} for the criteria for
inclusion in the Nominal sample.}

\tablenotetext{a}{Percent galaxy contamination, determined from the 
synthetic photometry of the spectra in the observer-frame $g$ band.}

\tablenotetext{b}{Phase relative to the time of $B$-band maximum light.}

\tablenotetext{c}{$M_{V} (t = 0) = -19.504 + 0.736 \Delta + 0.182
\Delta^{2} + 5 \log (H_{0}/65 {\rm ~km~s}^{-1}{\rm ~Mpc}^{-1})$~mag
\citep{Jha07}.}

\tablenotetext{d}{Determined from MLCS fits to our light curves.}

\tablenotetext{e}{Determined from \texttt{kcorrect} fits to the
host-galaxy broad-band photometry.}

\tablenotetext{f}{Percentage of the total star formation over
the last 1~Gyr compared to the total star formation over all time,
determined from \texttt{kcorrect} fits to the host-galaxy broad-band
photometry.}

\end{deluxetable*}
\end{center}

Although all SNe in our sample are obviously SNe~Ia based on spectral
features \citep[e.g.,][]{Filippenko97}, we have also performed the
Supernova Identification (SNID) procedure of \citet{Blondin07} for
each spectrum, confirming our classifications.  Heliocentric redshifts
were determined through either template matching (with SNID) or from
host-galaxy emission or absorption lines, with the precision depending
on the method (template matching vs.\ narrow host-galaxy lines) and
the resolution of the spectrum from which the redshift is measured.
Redshifts determined from template matching, from narrow lines in the
Keck spectra, and narrow lines in either the SDSS main sample or
Magellan host-galaxy spectra correspond to redshifts with 2, 3, and 4
significant digits in Table~\ref{t:spec} and uncertainties of \about
0.01, 0.001, and 0.0001, respectively.

The spectra are typically of higher quality than most of those
previously published for high-redshift SNe~Ia.  The single exception
is SN~16789, which has a relatively low S/N and has been removed from
our final sample for analysis (but its host will be discussed later).
There is one outlier (compared to the rest of the sample) in both
redshift and light-curve shape: SN~7147.  It is the lowest redshift SN
($z = 0.111$), and separated in redshift space from the rest of the
sample (the next lowest redshift is 0.180).  It also has a distinct
light curve; unlike the rest of the sample, which clusters near
$\Delta = -0.2$, SN~7147 has $\Delta = 0.36$, corresponding to a
relatively fast-declining light curve and indicating a
lower-luminosity SN.  Its spectrum also supports a lower luminosity,
with a ``silicon ratio'' of $\mathcal{R} ({\rm Si}) = 0.4$,
corresponding to lower-luminosity ``Branch-normal SNe~Ia''
\citep{Nugent95}.

\subsection{Galaxy Subtraction}\label{ss:galsub}

As seen in \citet{Foley08:comp}, despite careful spectral reductions
which attempt to remove host-galaxy light, many high-redshift SN~Ia
spectra have significant host-galaxy contamination.  Consequently, the
intrinsic SN SED is difficult to recover from the spectrum.  However,
since the parameter space of galaxy SEDs is well known (and well
behaved), one can reliably reconstruct galaxy SEDs with broad-band
photometry \citep[e.g.,][]{Blanton03}.

Adopting the approach described by \citet{Blanton03}, but updated to
include UV wavelengths by \citet{Blanton07}, and implemented in the
IDL software package
\texttt{kcorrect.v4\_1\_4}\footnote{http://howdy.physics.nyu.edu/index.php/Kcorrect
.}, we have used the \ugriz photometry of the host galaxy at the
position of the SN, which is a byproduct of the SMP method
\citep{Holtzman08}, and the redshifts presented in Table~\ref{t:spec}
to reconstruct, using a $\chi^{2}$ minimization method, the galaxy
SEDs at the position of the SN.  The host photometry used to
reconstruct the galaxy SEDs is therefore derived from the combination
of many Stripe 82 images.  Additionally, we employed the SDSS DR6
\citep{Adelman-McCarthy08} \ugriz photometry for the host galaxies to
determine galactic properties such as star-formation rate and mass.
Because of the filter gap between the $g$ and $r$ bands, reconstructed
SEDs of galaxies in the redshift range $0.27 < z < 0.33$, in which the
4000~\AA\ break is between the filters, are slightly degenerate.  This
degeneracy results in many SEDs with similar $g-r$ colors but vastly
different $u-g$, $r-i$, and $r-z$ colors; hence, this method can
create unphysical SEDs for these galaxies.  However, for the typical
galaxy with $z < 0.5$, the SEDs are recovered to $\lsim 0.02$~mag in
all filters \citep{Blanton03, Blanton07}.

Since there is more variability in the UV SEDs of galaxies than in
optical SEDs, the UV region of the SED is more difficult to
reconstruct than the optical region.  \citet{Blanton07} tested the
predictive properties of \texttt{kcorrect} by determining the SEDs of
galaxies with both {\it GALEX} and SDSS photometry using only the SDSS
photometry.  They then compared the residual of the reconstructed {\it
GALEX}$-$SDSS colors compared to the observed values.  For these
galaxies the mean of the residuals was very close to zero.  Since we
only wish to reconstruct the SED to \about 3000~\AA\ in the observer
frame, we are confident that the \ugriz\ data are sufficient for
reconstructing SEDs over the wavelength region of interest.
Additionally, since our SNe are found in Stripe 82, our host
photometry is typically much deeper than that of most SDSS galaxies.

\begin{figure}
\begin{center}
\epsscale{2.4}
\rotatebox{90}{
\plotone{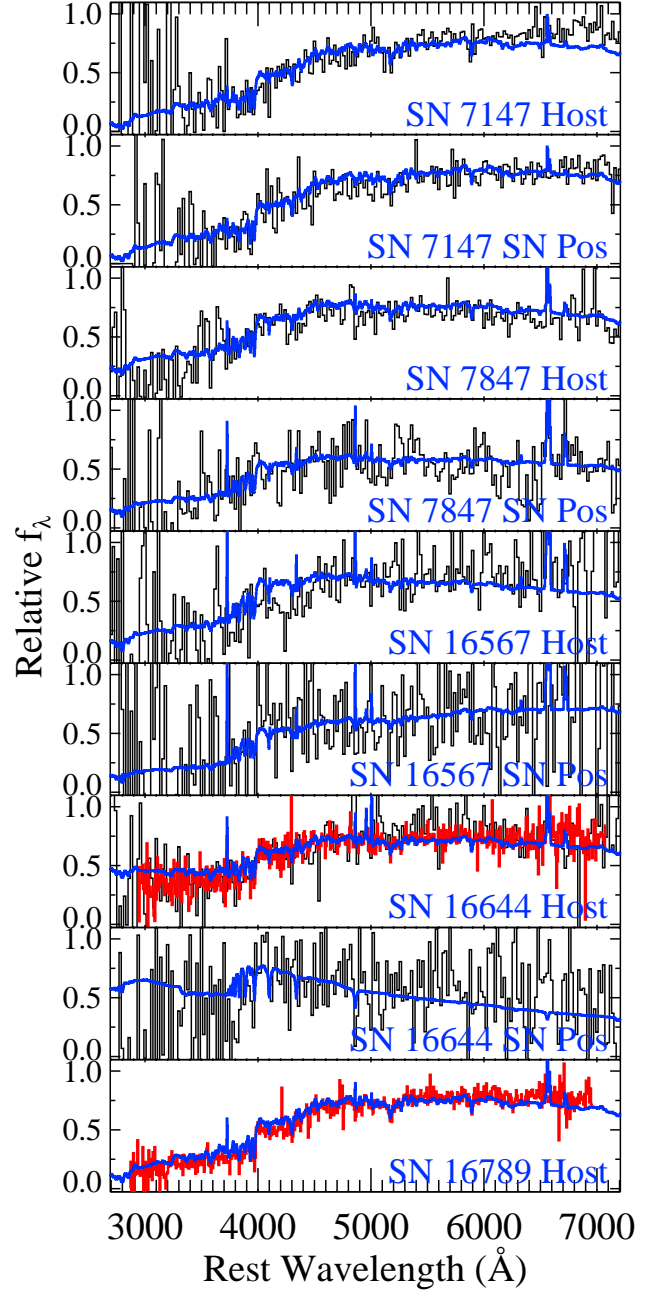}}
\caption{Magellan/MagE spectra (black curves) of the nucleus of
SDSS~J232004.44-000320.1 (the host galaxy of SN~7147; top panel),
SDSS~J232004.44-000320.1 at the position of SN~7147 (second panel),
the nucleus of SDSS~J020950.32-000342.0 (the host galaxy of SN~7847;
third panel), SDSS~J020950.32-000342.0 at the position of SN~7847
(fourth panel), the nucleus of SDSS~J014145.08-001121.8 (the host
galaxy of SN~16567; fifth panel), SDSS~J014145.08-001121.8 at the
position of SN~16567 (sixth panel), the nucleus of
SDSS~J022716.17-002336.5 (the host galaxy of SN~16644; seventh panel),
SDSS~J022716.17-002336.5 at the position of SN~16644 (eighth panel),
and the nucleus of SDSS~J025508.24+001405.2 (the host galaxy of
SN~16789; bottom panel).  The blue curves are the reconstructed galaxy
spectra determined by the \texttt{kcorrect} routine.  The red curves
are SDSS spectra of the galaxies.  It is worth noting how well the
reconstructed spectra fit the observed spectra over the majority of
optical wavelengths.}\label{f:gal_fit}
\end{center}
\end{figure}

Fortuitously, the hosts of SNe~16644 and 16789 have been observed
spectroscopically by the SDSS survey \citep{Adelman-McCarthy08},
allowing a check of the accuracy of this method for reconstructing the
galaxy spectra.  We also observed five host galaxies with the Magellan
Clay telescope after the SN had faded, where we were able to extract a
spectrum at the SN position and another spectrum at the position of
the host nucleus.  Figure~\ref{f:gal_fit} compares the observed and
reconstructed spectra of the hosts of SNe~7147, 7847, 16567, 16644,
and 16789.  For the spectra corresponding to the nucleus (including
the SDSS spectra), we present the reconstructed spectra of the entire
galaxy, and not at the position of the SN.  For the spectra
corresponding to the SN position, we present the reconstructed spectra
at that position.

The data agree very well over most of the wavelengths covered by
the observed spectra, suggesting that this procedure is a robust way
to determine host-galaxy SEDs.  Since SNe~16644 and 16789 are at $z =
0.2988$ and 0.3250, respectively, we can use these observations to
understand the potential systematic biases for objects with $0.27 < z
< 0.33$.  For most spectra of both the host nucleus and SN position,
the observed spectra and reconstructed spectra match very well.
However, there are slight differences for SN~16644.  The reconstructed
spectrum has more near-UV flux at both the nucleus and SN position.
Our sample contains eight SNe~Ia with $0.27 < z < 0.33$.  However,
only five of these SNe are used in our final analysis.  We have also
performed our analysis both with and without these SNe (see
Section~\ref{ss:composite}), and there is no significant difference in
the results.

After generating a reliable host-galaxy SED with the \texttt{kcorrect}
routine, we perform two separate methods for determining the SN SED.
The methods are different in approach, but yield similar results.
Both methods are described in detail in Appendix~\ref{a:galsub}.
Briefly, ``photometry matching'' is similar to the method employed by
\citet{Ellis08}.  It requires extracting the SN without attempting to
perform local galaxy subtraction and warping the resulting spectrum to
the combined host and SN photometry for an aperture corresponding to
the combination of slit and seeing.  The second method, ``color
matching,'' is a novel way to remove galaxy contamination by matching
the galaxy-subtracted spectrum to the SN colors at the time the
spectrum was obtained.  For our sample, both methods yielded similar
results.  Both methods have their advantages, but the color-matching
method, which removes as much galaxy contamination as possible at the
time of extraction, should have a smaller potential for introducing
systematic errors in our sample.  We have performed our analysis using
both methods, and the results do not change significantly.  We will
thus focus on the color-matched galaxy-subtracted spectra.

\subsection{Composite Spectra}\label{ss:composite}

The SN spectra presented in this paper are generally of very high
quality, allowing for studies of individual SNe.  However, much
additional information can be determined from a composite spectrum of
all SNe.  We are able to examine both the average properties of the
sample as well as small-amplitude spectral features which are below
the noise level in individual spectra \citep{Foley08:comp, Ellis08,
Sullivan09}.

Using the same procedure to produce the composite spectra as in
\citet{Foley08:comp}, we created several composite spectra from
our sample (details of the sample are presented by
\citealt{Foley08:comp} and Silverman et al.; submitted)).  This
includes removing residual emission lines and normalizing the spectra
in the region $4500 \le \lambda \le 7500$~\AA.  The properties of each
composite spectrum are listed in Table~\ref{t:comp}.

\begin{center}
\begin{deluxetable*}{l@{ }r@{ }c@{ }c@{ }c@{ }c@{ }c@{ }c@{ }c@{ }c@{ }c}
\tabletypesize{\scriptsize}
\tablewidth{0pt}
\tablecaption{Composite Spectra Information\label{t:comp}}
\tablehead{
\colhead{} &
\colhead{} &
\multicolumn{5}{c}{Supernovae Included in Subsample?} &
\colhead{} &
\colhead{} &
\colhead{} &
\colhead{UV Flux} \\
\colhead{Composite} &
\colhead{No. of} &
\colhead{} &
\colhead{High-} &
\colhead{} &
\colhead{} &
\colhead{} &
\colhead{Mean} &
\colhead{Mean} &
\colhead{Mean} &
\colhead{Relative to} \\
\colhead{Name} &
\colhead{Spectra} &
\colhead{Rejected} &
\colhead{Contamination} &
\colhead{$0.27 < z < 0.33$} &
\colhead{$|t| > 3$~days} &
\colhead{SN~7147} &
\colhead{Phase (days)} &
\colhead{Redshift} &
\colhead{$\Delta$} &
\colhead{Nominal}}

\startdata

All               & 21 & Y & Y & Y & Y & Y & $-0.4$  & 0.23 & $-0.15$ & 0.99 \err{0.07}{0.06} \\
Nominal           & 17 & N & Y & Y & Y & Y & $-0.3$  & 0.22 & $-0.14$ & \nodata \\
Maximum           & 10 & N & Y & Y & N & Y & \phs0.4 & 0.22 & $-0.12$ & 1.01 \err{0.09}{0.09}  \\
Low Galaxy        & 11 & N & N & Y & Y & N & \phs0.0 & 0.23 & $-0.16$ & 1.01 \err{0.07}{0.07}  \\
Low $\Delta$      & 16 & N & Y & Y & Y & N & $-0.4$  & 0.23 & $-0.18$ & 1.01 \err{0.07}{0.06}  \\
No Gap            &  6 & N & N & N & Y & N & $-1.1$  & 0.19 & $-0.17$ & 1.20 \err{0.05}{0.27}  \\
Strict            &  3 & N & N & N & N & N & \phs0.4 & 0.20 & $-0.16$ & 1.22 \err{0.11}{0.14}  \\
\hline										             
\multicolumn{10}{c}{Nominal Subsamples} \\					             
\hline										             
Active            &  7 & N & Y & Y & Y & N & $-1.2$  & 0.24 & $-0.22$ & 1.02 \err{0.09}{0.11} \\
Passive           &  6 & N & Y & Y & Y & Y & \phs1.0 & 0.19 & $-0.03$ & 1.00 \err{0.14}{0.15} \\
Low Mass          & 10 & N & Y & Y & Y & N & $-1.4$  & 0.23 & $-0.18$ & 0.99 \err{0.07}{0.07} \\
High Mass         &  5 & N & Y & Y & Y & Y & \phs0.9 & 0.18 & $-0.02$ & 1.00 \err{0.16}{0.16} \\
Lower $z$         &  9 & N & Y & N & Y & Y & $-1.3$  & 0.19 & $-0.12$ & 1.02 \err{0.09}{0.09} \\
Higher $z$        &  8 & N & Y & Y & Y & N & \phs2.2 & 0.31 & $-0.19$ & 0.95 \err{0.07}{0.06} \\
Negative $\Delta$ &  8 & N & Y & Y & Y & N & $-1.2$  & 0.24 & $-0.28$ & 1.07 \err{0.08}{0.10} \\
Zero $\Delta$     &  9 & N & Y & Y & Y & Y & \phs0.4 & 0.21 & $-0.02$ & 0.94 \err{0.10}{0.08}

\enddata

\end{deluxetable*}
\end{center}

The various composite spectra are composed of subsamples of our main
sample, culled by quality cuts.  The ``All'' spectrum contains the
entire sample.  All other composite spectra exclude the SNe which fail
the galaxy-contamination correction.  The ``Nominal'' sample includes
all SNe except those that fail the galaxy-contamination routine.
Since SN~Ia spectra evolve with time, having a small time window will
decrease discrepancies that are the result of large differences in the
epoch of spectra.  Accordingly, we created a ``Maximum''
composite spectrum removing the SNe with $|t| > 3$~days (relative to
rest-frame $B$ maximum brightness) from the Nominal sample.  To reduce
sensitivity to host-galaxy contamination, we created the ``Low
Galaxy'' composite spectrum, removing the SNe with high galaxy
contamination ($> 8\%$ in the $g$ band) from the Nominal sample.
Finally, we created a ``Low-$\Delta$'' composite spectrum, removing
the low-redshift, high-$\Delta$ SN~7147, which might not be
representative of the high-redshift population of SNe~Ia and could
bias our sample, from the Nominal sample.

We have further considered the effect of having SNe with $0.27 < z <
0.33$ in our sample.  Although these SNe pass our galaxy-subtraction
routines, they may have incorrect galaxy subtraction.  We have removed
these SNe from the Low-Galaxy sample to create the ``No Gap'' sample.

\begin{figure}
\begin{center}
\epsscale{2.73}
\rotatebox{90}{
\plotone{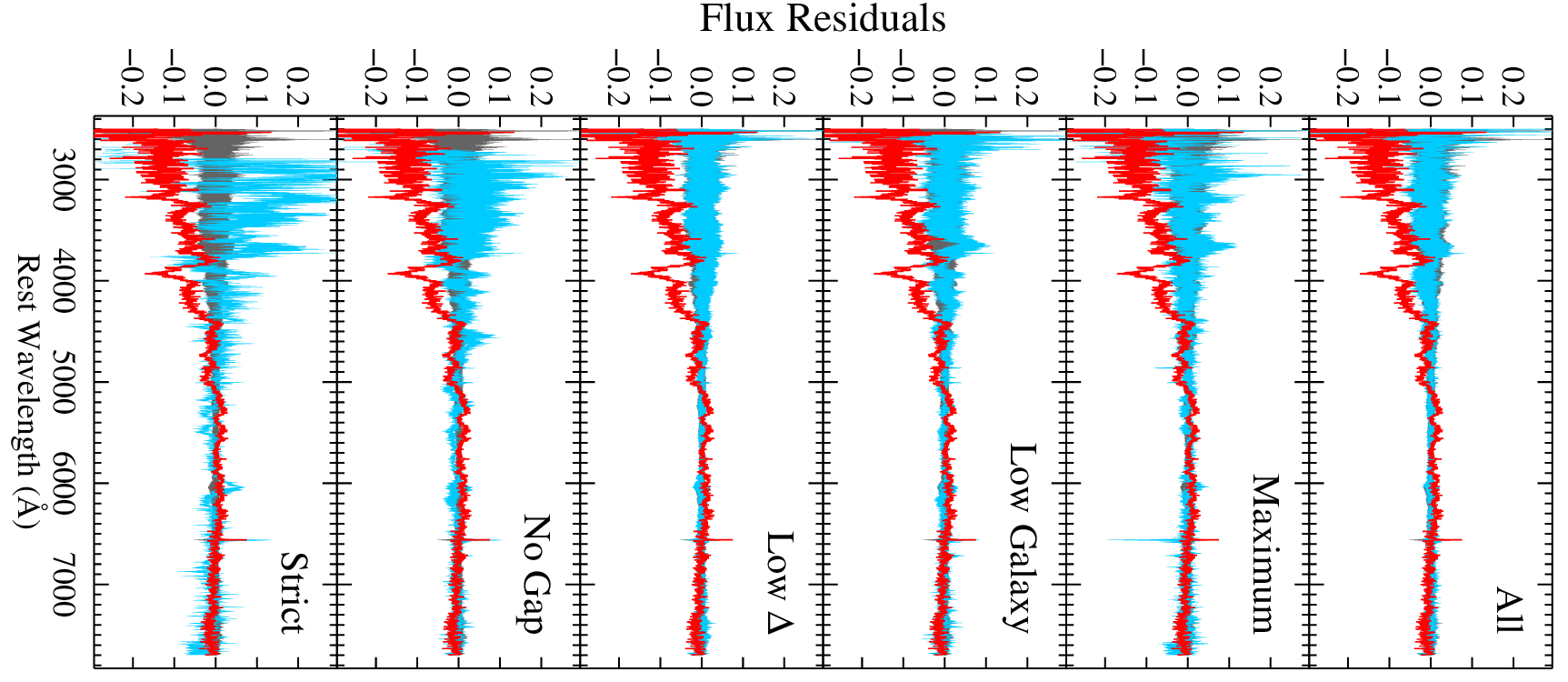}}
\caption{The 1$\sigma$ boot-strap sampling error region for the Nominal
composite spectrum is displayed in grey.  The 1$\sigma$ boot-strap
sampling error residuals for the (from top to bottom) All, Maximum,
Low Galaxy, Low $\Delta$, No Gap, and Strict composite spectra
compared to the Nominal composite spectrum are in light blue (X minus
Nominal).  The red curve is the residual of the Lick composite and
Nominal Keck/SDSS composite spectra (Lick minus Nominal).  The grey
region and red curves are the same in each panel.  The differences
between some spectra at \about 3650~\AA\ and \about 6150~\AA\ are the
result of slightly different mean phases, causing a shift in the
velocity of the Ca H\&K and \ion{Si}{2} $\lambda$6355 features,
respectively (see
Section~\ref{ss:sample_selection}).}\label{f:comp_diff}
\end{center}
\end{figure}

To determine how consistent each composite spectrum is with the
Nominal composite spectrum in the UV, we have determined the ratio of
the UV flux in the wavelength region 2700--4000~\AA\ for each
composite spectrum relative to the Nominal composite spectrum and
listed the value in Table~\ref{t:comp}.  The uncertainties given in
the table are determined from the 1$\sigma$ boot-strap sampling error
spectrum relative to the Nominal composite spectrum.  These
uncertainty measurements do not include the boot-strap sampling errors
of the Nominal spectrum, but we note that the 1$\sigma$ boot-strap
sampling error spectrum for the Nominal sample results in a flux that
is 0.94 or 1.06 (lower and upper bounds) times the flux of the Nominal
composite spectrum in this region.  This metric does not contain all
pertinent information in the spectrum.  For instance, two spectra can
have the same average flux over a wavelength range but have
significantly discrepant spectral features over that wavelength range.
Therefore, we caution use of this number to determine the consistency
of two composite spectra, but rather to use these values as indicators
of differences while examining the composite spectra (i.e.,
Figure~\ref{f:comp_diff}) for a detailed comparison.

Figure~\ref{f:comp_diff} shows that to within the uncertainties, the
Nominal composite spectrum is similar to all other composite spectra
mentioned so far, indicating that the cuts to the Nominal sample do
not affect our results, and that the Nominal sample is not
significantly biased.  Finally, the ``Strict'' composite spectrum
excludes all SNe in at least one of the following categories: fail the
galaxy-contamination routine, have $|t| > 3$~days, have high galaxy
contamination, SN~7147, and have $0.27 < z < 0.33$.  The Strict sample
consists of just three SNe, and although they do not have any obvious
deficiencies, the small sample size means that a single peculiar
object may significantly bias the sample.

Examining Figure~\ref{f:comp_diff} in detail, we see that the
Low-Galaxy, No Gap, and Strict composite spectra differ slightly from
the Nominal composite spectrum.  The differences between the
Low-Galaxy and Nominal composite spectra are confined to the regions
of the spectrum corresponding to the Ca H\&K feature, and are likely
the result of slightly discrepant line velocities or line strengths.
The differences in line velocities are almost certainly the result of
having slightly different mean phases (e.g., \citealt{Foley11:vgrad};
see also Section~\ref{ss:sample_selection}).  The UV flux for the
Low-Galaxy composite spectrum is 1.01 \err{0.07}{0.07}, further
indicating that the differences are confined to a spectral feature and
do not change the continuum.  The No Gap and Strict composite spectra
have more UV flux than the Nominal composite spectrum (having UV flux
of 1.20 \err{0.05}{0.27} and 1.22 \err{0.11}{0.14} times that of the
Nominal composite spectrum, respectively).  The No Gap spectrum has
slightly more UV flux than the Nominal spectrum, but the spectra are
consistent within the boot-strap sampling errors.  The Strict spectrum
has a larger deviation from the Nominal spectrum, but the Strict
spectrum consists of only three SN spectra and may be biased.  As
shown in Section~\ref{s:comp}, the Nominal spectrum has more UV flux
than the low-redshift composite spectrum; the latter has a UV flux
0.84 \err{0.07}{0.06} times that of the Nominal composite spectrum.
The No Gap and Strict spectra have larger deviations with the
low-redshift composite spectrum, indicating that quality cuts cannot
account for the difference.

We are confident that the Nominal spectrum is an excellent
representative of this sample of SNe~Ia.  No significant differences
are observed when restricting our sample to just the SNe near maximum
brightness or having low galaxy contamination.  There is a minor, but
insignificant, difference in the UV continuum of the Nominal and No
Gap composite spectra.  The No Gap sample contains only six SNe, and
is therefore more susceptible to biases.  Similarly, the Strict
composite spectrum, which also deviates slightly from the Nominal
composite spectrum at bluer wavelengths, is derived from only three
SNe.  Given the composite spectra generated from various subsamples
and quality-cut tests we have performed, we see no significant biases
or systematic errors associated with the Nominal sample and its
composite spectrum.  Thus, we will use the Nominal spectrum for our
analysis.


\section{Comparison of the Keck/SDSS with Other Supernova Samples}\label{s:comp}

\subsection{Composite Spectra}\label{ss:comp}

\begin{figure*}
\begin{center}
\epsscale{1.45}
\rotatebox{90}{
\plotone{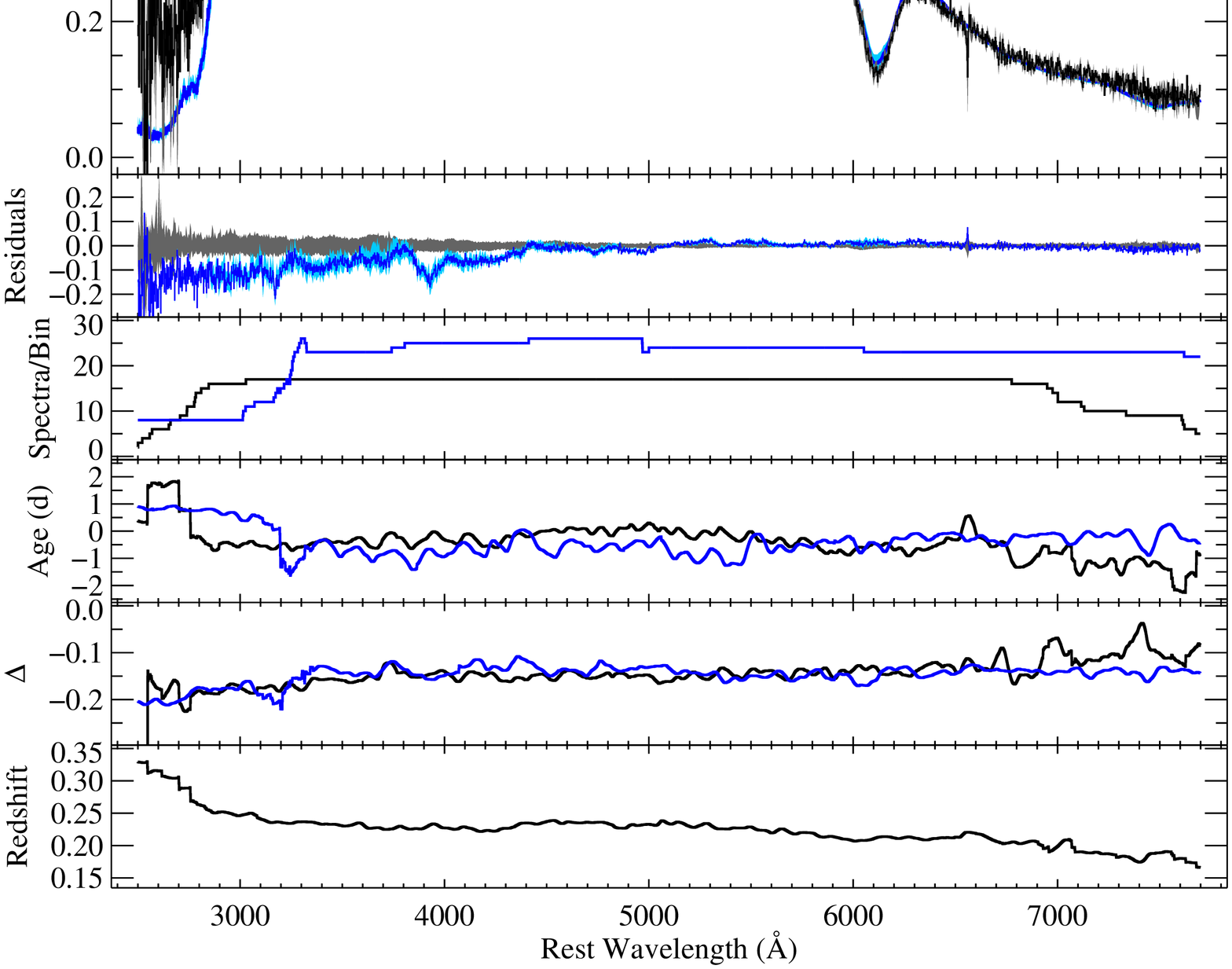}}
\caption{({\it top panel}):  Composite spectrum created from our
Keck/SDSS sample (black curve) compared to the maximum-light
low-redshift composite spectrum (blue curve).  The spectra are scaled
to match in the region $4500 \le \lambda \le 7500$~\AA.  The grey and
light-blue regions are the 1$\sigma$ boot-strap sampling errors for
the Keck/SDSS and low-redshift composite spectra, respectively.  ({\it
second panel}): The grey region is the 1$\sigma$ boot-strap sampling
region for the Keck/SDSS composite spectrum.  The blue curve is the
residual of the Keck/SDSS and low-redshift composite spectra.  The
light-blue region is the residual of the Keck/SDSS composite spectrum
and the low-redshift 1$\sigma$ boot-strap sampling region.  ({\it
third panel}): The number of individual spectra contributing to each
wavelength bin in the Keck/SDSS (black curve) and low-redshift (blue
curve) composite spectra.  ({\it fourth panel}): The average phase
relative to maximum brightness as a function of wavelength for the
Keck/SDSS (black curve) and low-redshift (blue curve) composite
spectra.  ({\it fifth panel}): The average value of $\Delta$ as a
function of wavelength for the Keck/SDSS (black curve) and
low-redshift (blue curve) composite spectra. ({\it bottom panel}): The
average redshift of the Keck/SDSS composite spectrum as a function of
wavelength.}\label{f:comp}
\end{center}
\end{figure*}

In Figure~\ref{f:comp}, we present the Nominal composite spectrum and
compare it with a low-redshift maximum-light composite spectrum.  This
low-redshift composite spectrum was constructed from the sample of
low-redshift spectra used by \citet{Foley08:comp}, with the additional
UV spectra of SN~2005cf \citep{Bufano09}.  This sample represents one
of the largest and cleanest sets of low-redshift SN~Ia spectra.  Most
spectra were obtained with the Lick 3~m telescope and Kast double
spectrograph \citep{Miller93}, which provides well-calibrated spectra
from the atmospheric limit to 1~$\mu$m (see \citealt{Foley08:comp} and
Silverman et~al., submitted, for additional details).  The
low-redshift sample used to construct the composite spectrum is of
very high quality.  The spectra are of well-monitored (both
photometrically and spectroscopically) SNe~Ia.  Each spectrum is
selected to have very low galaxy contamination (typically $< 3\%$) and
very high S/N (typically $> 50$).  The phase and $\Delta$ range for
the low-redshift sample were chosen to match those of the Keck/SDSS
sample.  To achieve this, the low-redshift spectra have $-4.8 < t <
3.0$~days and $\Delta < 0.1$.  From the fourth and fifth panels of
Figure~\ref{f:comp}, it is clear that the average values for phase and
$\Delta$ are very similar between the two samples at all wavelengths.

For wavelengths longer than \about 5000~\AA, the Keck/SDSS and
low-redshift composite spectra are remarkably similar.  At wavelengths
shorter than \about 4400~\AA, however, the spectra differ; {\it the
Keck/SDSS spectrum is bluer than its low-redshift counterpart}.  The
difference between the spectra grows larger with decreasing
wavelength, becoming more significant by 3000~\AA.  For wavelengths $<
3400$~\AA, the Keck/SDSS composite spectrum has $>20$\% more flux than
the low-redshift composite spectrum.  Despite the differences in
overall spectral shape, the individual features are all consistent
between the two composite spectra.

As mentioned above, it is important to note that the average
properties of the spectra (namely redshift, phase, and $\Delta$)
contributing to a given wavelength bin can change with wavelength.
This is particularly the case for the spectra covering slightly
different rest-frame wavelength ranges.  For the Keck/SDSS sample, the
change of average values with wavelength is the result of different
redshifts for the SNe, while the average values change in the
low-redshift sample mainly because of the different samples for
optical and UV spectra.

Considering how well matched the Keck/SDSS and low-redshift composite
spectra are in terms of phase and $\Delta$ (the two parameters known
to affect an SN spectrum), the differences must come from a different
source of variation.  Specifically, the samples may be mismatched in
some other parameter not known to cause significant color differences,
or there is some uncharacterized problem with our data reduction
techniques.  Below we examine additional parameters and their effects
on the spectra.

\subsection{Supernova Colors}\label{ss:colors}

The high quality of our spectra presents the opportunity to derive
properties of the individual SNe.  It was shown in
Section~\ref{ss:comp} that the Keck/SDSS composite spectrum is bluer
than the low-redshift composite spectrum at UV wavelengths.  Although
comparison of the low-redshift composite spectrum to each individual
Keck/SDSS SN spectrum is possible, the S/N is much lower in the
individual spectra.  However, by binning the spectra, we may be able
to sufficiently increase the S/N of the individual spectra to make an
interesting measurement.

\begin{figure}
\begin{center}
\epsscale{0.85}
\rotatebox{90}{
\plotone{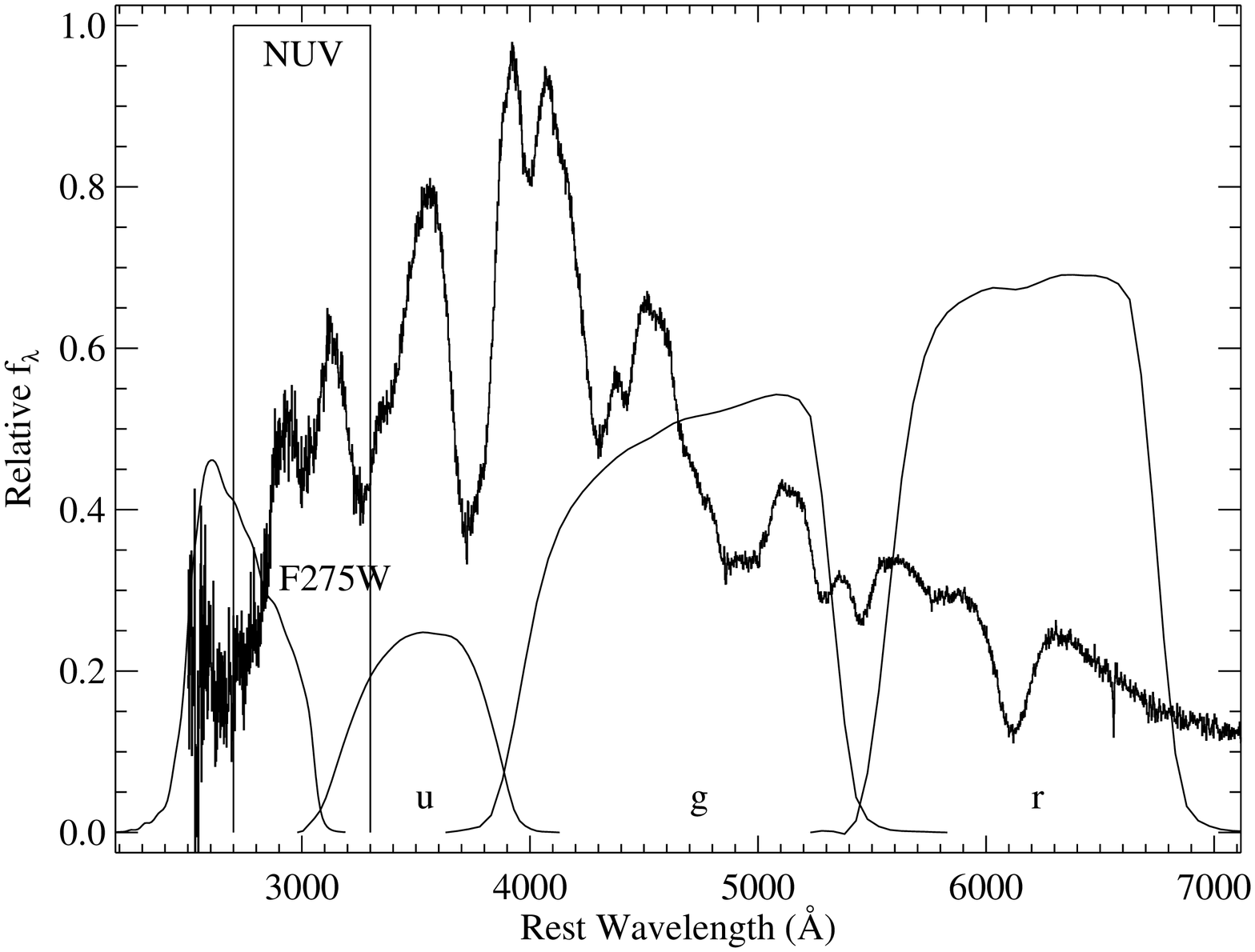}}
\caption{The composite Keck/SDSS spectrum shown in Figure~\ref{f:comp}.
Overplotted are (from blue to red) the transmission curves of the {\it
HST} WFC3 F275W, the newly defined NUV, and the SDSS \ugr filters.
The noise in the Keck/SDSS composite spectrum for rest wavelengths
shorter than 2500~\AA\ is substantial and not
plotted.}\label{f:filters}
\end{center}
\end{figure}

The binning that we choose is to convolve the spectra with various
broad-band filters.  Figure~\ref{f:filters} displays the Keck/SDSS
composite spectrum with the transmission curves of the {\it HST} WFC3
F275W and Keck/SDSS \ugr filters.  We also introduce the ``NUV''
(near-UV) filter, a top-hat filter that spans the wavelength range
2700--3300~\AA.  The $g$ and $r$ filters probe the area of the
spectrum that is mostly consistent between the low-redshift and
Keck/SDSS composite spectra; however, the blue edge of the $g$ band
does show some differences, so the $g-r$ color of the Keck/SDSS
composite spectrum should be bluer than that of the low-redshift
composite spectrum.  Meanwhile, the F275W, NUV, and $u$ filters probe
the region where the spectra diverge.

\begin{figure}
\begin{center}
\epsscale{1.45}
\rotatebox{90}{
\plotone{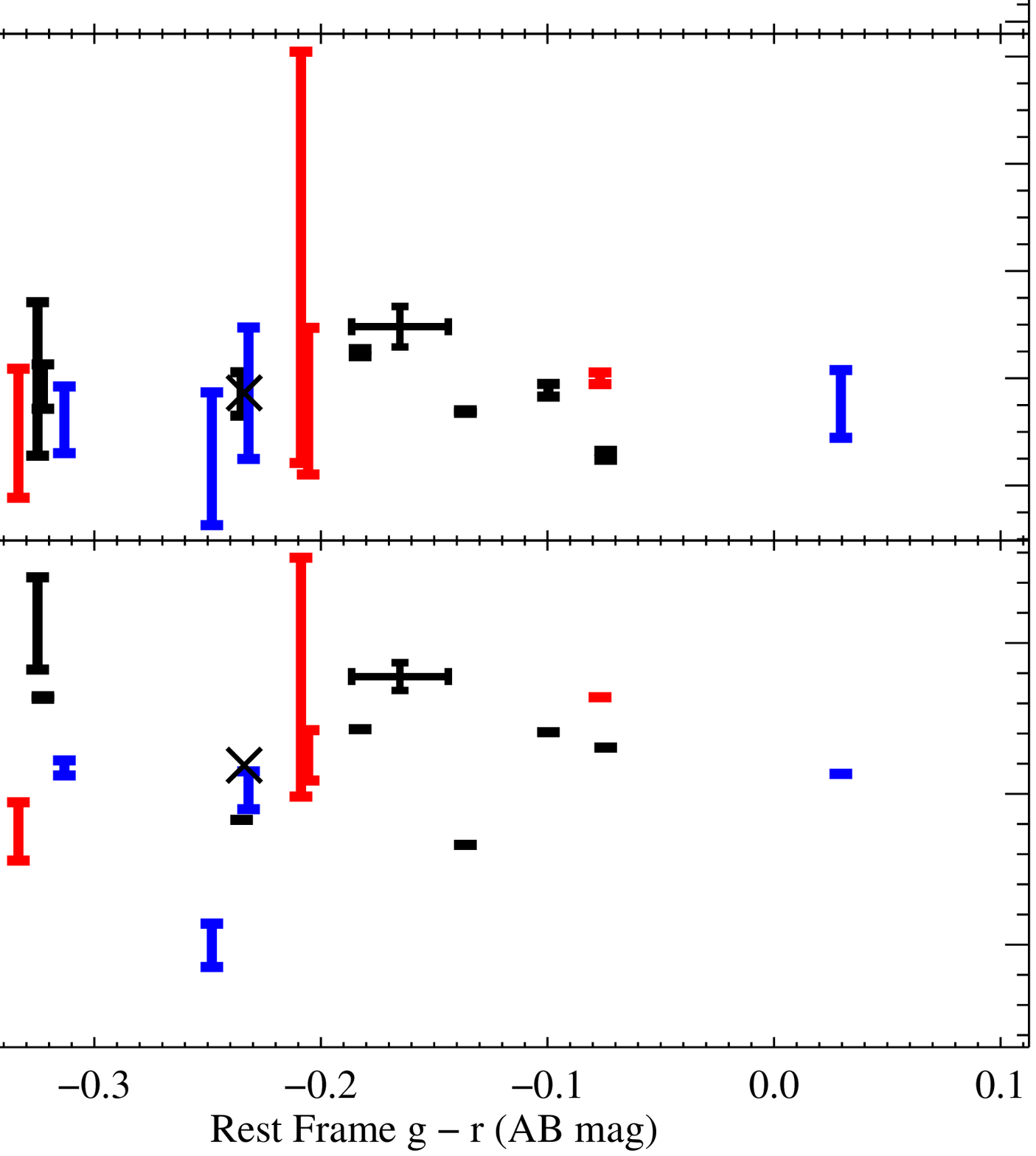}}
\caption{Color-color diagrams for the individual Keck/SDSS SNe~Ia.
The black, blue, and red points correspond to SNe with no, low ($<
8\%$), and high ($> 8\%$) host-galaxy contamination, respectively.
The black cross with error bars in both directions indicates the
colors measured for the low-redshift composite spectrum.  The black
``X'' indicates the Keck/SDSS composite spectrum.  The color
measurements are derived by convolving the individual Keck/SDSS SN
spectra with filter transmission curves.  Since not all of the spectra
cover the entire wavelength range of each filter, there is a range of
color for some points.  The upper limits are derived by setting all
flux for wavelengths shorter than the shortest observed wavelength of
a spectrum to zero.  The lower limits are derived by setting all flux
for wavelengths shorter than the shortest observed wavelength of a
spectrum to the median flux value of the bluest 100~\AA\ of the
spectrum.}\label{f:spec_colors}
\end{center}
\end{figure}

Using the filter transmission functions, we made color measurements
for all of the Keck/SDSS spectra.  We have done the same for the
low-redshift composite spectrum.  Unfortunately, the low-redshift
sample rarely has concurrent UV and optical spectra for individual
SNe; we therefore are unable to measure the UV minus optical colors of
individual SNe.  To produce reasonable uncertainties for the Lick
composite spectrum, we performed a Monte Carlo analysis of the
spectrum using the boot-strap sampling errors to produce several
hundred SN spectra.  Our measured uncertainties are the standard
deviation of the colors for these spectra.  The color measurements are
presented in Figure~\ref{f:spec_colors}.

Some of our SNe have rest-frame spectra that do not extend to
sufficiently short wavelengths to cover the entire filter. We have
taken two approaches to measure synthetic colors for these spectra.
First, we set all fluxes for wavelengths shorter than the minimum
observed wavelength in the spectrum to zero.  For SNe with spectra
that extend blueward of 2500~\AA, where the spectra can be
particularly noisy, we also set the flux to zero.  Since any flux at
these wavelengths will create a bluer color, this measurement sets an
upper limit for the color.  The second approach sets the fluxes for
wavelengths shorter than the minimum observed wavelength in the
spectrum to the median flux value for the bluest 100~\AA\ of the
spectrum.  Since the flux generally increases with wavelength in the
UV, this is likely an overestimate of the flux and sets a reasonable
lower limit for the color.

As expected from the comparison of the composite spectra, the
Keck/SDSS SNe have $g-r$ colors that are similar to those of the
low-redshift composite spectrum, with the composite spectra differing
by $0.069 \pm 0.025$~mag.  Nearly all of the Keck/SDSS SNe have bluer
${\rm F275W} - g$ and ${\rm NUV} - g$ colors than the low-redshift
composite spectrum, with the composite spectra differing by 0.618 and
0.294~mag for those colors, respectively.  Similarly, most of the
Keck/SDSS SNe have bluer $u - g$ colors than the low-redshift
composite.  Using the Student's $t$-test and assuming that the Lick
composite spectrum has the correct colors, we find that the
probabilities of the Keck/SDSS and low-redshift samples having the
same mean colors are $9.4 \times 10^{-6}$, $1.2 \times 10^{-4}$, and
$1.5 \times 10^{-2}$ (corresponding to 4.4$\sigma$, 3.8$\sigma$, and
2.4$\sigma$) for the ${\rm F275W} - g$, ${\rm NUV} - g$, and $u - g$
colors, respectively.  Conversely, we find that for the $g - r$ color,
the two samples are completely consistent with being drawn from the
same population.

The color information from the individual spectra shows that the bluer
color of the Keck/SDSS composite spectrum is not the result of a
minority of SNe heavily influencing the overall spectral shape.  The
probabilities also indicate that it is unlikely that the low-redshift
and Keck/SDSS samples have similar UV colors.

\section{Implications for Supernova Cosmology}\label{s:cosmo}

In Section~\ref{s:comp}, we have shown that the low-redshift and
Keck/SDSS samples appear to have different UV SEDs.  We will examine
possible explanations for these differences, including systematic
effects, in Section~\ref{s:disc}.  However, for now, we examine some
implications of such a difference for cosmological results.  In
particular, we explore whether such a mismatch in SEDs could produce
the $U$-band anomaly.

As discussed in Section~\ref{s:intro}, \citetalias{Kessler09:cosmo}
found differences in the rest-frame $U$-band data for low-redshift and
high-redshift samples that, depending on the treatment of the
rest-frame $U$ band, resulted in a systematic shift in the measured
value of $w$.  This ``$U$-band anomaly'' is larger for some datasets
than others.

We first examine single-band UV distance measurements in
Section~\ref{ss:singleband}, finding that a UV excess produces a
distance bias that is in the opposite direction to that of the
$U$-band anomaly.  In Section~\ref{ss:multiband}, we find that
differences between the observed and model UV SEDs significantly
affect the relationship between light-curve shape and observed color.
This explains the relationship between the UV flux deficit found by
\citetalias{Kessler09:cosmo} and the distance modulus offset.

\subsection{Single-Band UV Distance Estimates}\label{ss:singleband}

We have shown that there is a mismatch in the UV SEDs of the
low-redshift and Keck/SDSS samples of SNe~Ia.  In this section, we
show that such a mismatch should affect the estimated distance to a
SN~Ia.  However, as we will show in Section~\ref{ss:multiband}, this
effect is dominated by effects related to the correlation between
light-curve shape and color when using multiple-band data, and is
likely not the cause of the $U$-band anomaly, but could be important
for some high-redshift SNe~Ia, which have limited color information.

Some light-curve fitters train on a low-redshift sample and apply the
measured relationships between light-curve shape and color from that
sample to high-redshift SNe; MLCS2k2 \citep{Jha07}, in particular,
uses this methodology.  Other light-curve fitters, such as SALT2
\citep{Guy07}, determine correlations between light-curve shape, color,
and peak luminosity using all SNe in its sample, including both low
and high-redshift data.

MLCS2k2 is trained on the low-redshift sample of SNe~Ia.  If the
average SEDs of low and high-redshift SNe~Ia in our observed samples
are properly described by the spectra presented in
Section~\ref{s:comp}, then MLCS2k2 will underpredict the flux of a
high-redshift SN~Ia in the rest-frame UV.

Imagine that we observed an SN~Ia at high redshift in only rest-frame
$U$.  The MLCS2k2 estimate of $M_{U, {\rm true}}$, the real peak
absolute magnitude of the SN in $U$, will be overestimated (MLCS2k2
will underestimate the luminosity of the SN in $U$).  We can define
this difference as
\begin{equation}
  M_{U, {\rm true}} = M_{U, {\rm MLCS}} - \Delta M,
\end{equation}
where $M_{U, {\rm MLCS}}$ is the MLCS2k2 estimate of the true value of
$M_{U, {\rm true}}$, and $\Delta M$ is the magnitude difference
between the estimate and the true value.  With our assumption and
formalism, $\Delta M \ge 0$~mag.

Since the observed magnitude, $U$, does not rely on any assumptions
about the SN SED, we find
\begin{align}
  \mu_{\rm true} &= U - M_{U, {\rm true}} = U - M_{U, {\rm MLCS}} + \Delta M \notag \\
                 &= \mu_{\rm MLCS} + \Delta M,
\end{align}
where $\mu_{\rm true}$ and $\mu_{\rm MLCS}$ are the true and
MLCS2k2-measured distance moduli, respectively.  Since $\Delta M \ge
0$~mag, MLCS2k2 is underestimating the distance to this SN.

\citetalias{Kessler09:cosmo} found that MLCS2k2 fits including and
excluding the rest-frame $U$ band produced a difference in average
distance moduli for the SDSS-II sample of $\mu_{{\rm no } U} - \mu
\simeq 0.1$--0.15~mag, where $\mu_{{\rm no } U}$ and $\mu$ are the
distance moduli measured with MLCS2k2 excluding and including the
rest-frame $U$ band, respectively.  If we assume that ignoring the
rest-frame $U$ band will avoid the complications of any potential SED
differences in the UV, then $\mu_{{\rm no } U}$ should be an unbiased
measurement of the true distance modulus, while $\mu$ represents the
measurement biased by the rest-frame $U$-band data.  Under this
assumption, $\mu_{{\rm no } U} = \mu_{\rm true}$ and $\mu = \mu_{\rm
MLCS}$.  Then we have
\begin{align}
  \Delta M &= \mu_{\rm true} - \mu_{\rm MLCS} \notag \\
           &= \mu_{{\rm no } U} - \mu \notag \\
           &\simeq 0.1 \mbox{--} 0.15 {\rm ~mag}.
\end{align}
Since $\Delta M > 0$~mag, the effect of the $U$-band anomaly is in the
same direction as that expected by an excess of UV flux in
high-redshift SNe~Ia relative to low-redshift SNe~Ia.

To determine if the amplitude of the effect is the same, we can simply
examine the $u-r$ color of the Keck/SDSS and low-redshift composite
spectra.  For this, we use the fact that the SEDs of both composite
spectra are very similar in the $r$ band; therefore, most differences
in this color will be from differences in the $u$ band.  In
Section~\ref{ss:colors}, we showed that the difference in $u-r$ colors
of the composite spectra is $\Delta (u-r) = 0.15 \pm 0.05$~mag.  This
difference is also of the same magnitude as seen by
\citetalias{Kessler09:cosmo}.

Another way in which a difference in UV SEDs can affect the measured
distance is the treatment of dust.  Since MLCS2k2 uses the SN colors
to estimate the host-galaxy extinction, a color difference not
accounted for by the model will result in an incorrect measurement of
this extinction.  (SALT2 does not attempt to directly measure
extinction, but the measured distance is still affected by SN color in
a similar manner.)

MLCS2k2 models all SNe~Ia with an intrinsic color (for our purposes,
we will use rest-frame $u-r$ as an example) that depends on
light-curve shape and has an intrinsic scatter.  If the measured color
is redder than the expected color for a given light-curve shape, then
MLCS2k2 models the difference in color as host-galaxy reddening.  If
high-redshift SNe~Ia have bluer $u-r$ colors than low-redshift SNe~Ia
(and our measurements of the composite spectra support this), then
MLCS2k2 will not properly measure the host-galaxy reddening,
estimating a value that is less than the true value.  Another way to
interpret this is that the MLCS2k2 zero-reddening color, $(u-r)_{0,
{\rm MLCS}}$, is too red.

We can describe this effect mathematically.  First, we define the real
extinction in the $V$ band as
\begin{equation}
  A_{V, {\rm true}} = \alpha \left [ (u-r) - (u-r)_{0, {\rm true}} \right ],
\end{equation}
where $\alpha$ is a factor that converts the color difference from the
zero-reddening value to extinction in the $V$ band (and is in part
related to $R_{V}$), $(u-r)$ is the measured color, and $(u-r)_{0,
{\rm true}}$ is the real zero-reddening color.  The above effect can
be characterized as
\begin{align}
  A_{V, {\rm MLCS}} &= \alpha \left [ (u-r) - (u-r)_{0, {\rm MLCS}} \right ] \notag \\
                    &= \alpha \left [ (u-r) - \left ((u-r)_{0, {\rm true}} + \Delta (u-r) \right ) \right ] \notag \\
                    &= A_{V, {\rm true}} - \alpha \left [ \Delta (u-r) \right ],
\end{align}
where $\Delta (u-r)$ is the difference between the MLCS2k2 and real
zero-reddening color.  Since the real zero-reddening color is bluer
than the MLCS2k2 zero-reddening color, $\Delta (u-r) \ge 0$~mag and
$A_{V, {\rm MLCS}} \le A_{V, {\rm true}}$.

Under these simple assumptions, we expect MLCS2k2 to underestimate the
extinction to a given SN~Ia, and therefore to underestimate its
luminosity and distance.  Since this effect is in the same direction
as the luminosity effect described above, the two effects should not
cancel each other out.

Although our simple analysis shows that a difference in the UV SED for
low and high-redshift SNe~Ia, similar to what is seen between the
low-redshift and Keck/SDSS samples, can produce a difference in the
inferred distances (both in amplitude and direction) as the $U$-band
anomaly, there is a significant observational problem: the SDSS-II
sample of \citetalias{Kessler09:cosmo} has, on average, lower $U$-band
flux than their adopted low-redshift sample.  This is opposite to the
difference seen in Figure~\ref{f:comp}.  We discuss the potential
reasons for the difference between the Keck/SDSS and overall SDSS-II
samples in Section~\ref{ss:sample_selection}.  Below, we discuss how
the full SDSS-II sample has the opposite UV flux difference (relative
to the low-redshift sample) yet affects the distance modulus in the
same direction as shown for the single-band case.

\subsection{Multi-Band UV Distance Estimates}\label{ss:multiband}

Our presentation of light-curve fitters in the previous section is
extremely simplified.  In reality, all light-curve data, including
bands other than rest-frame $U$ and data obtained away from maximum
light, affect the final estimated distance.  Although the difference
in the SEDs results in a distance modulus difference that is of the
same magnitude of the $U$-band anomaly, \citetalias{Kessler09:cosmo}
calculated distance moduli using three filters, and naively, one would
expect the rest-frame $U$ band to contribute about one third of any
difference.

Additionally, colors, which are only measured with multi-band data,
can influence the inferred distance modulus through the correlations
between observed color, light-curve shape, and peak luminosity.  In
the MLCS framework, an SN that is bluer than the SN model will be
assigned a $\Delta$ (the MLCS2k2 light-curve shape parameter) that is
smaller than it should, and the SN is assumed to be intrinsically
brighter than it is.  That is,
\begin{equation}
  M_{\rm true} > M_{\rm MLCS}.
\end{equation}
From this, we see that the distance modulus is affected such that
\begin{align}
  \mu_{\rm true} &= m - M_{\rm true} \notag \\
                 &< m - M_{\rm MLCS} \notag \\
                 &= \mu_{\rm MLCS}.
\end{align}
Therefore,
\begin{equation}
  \Delta M = \mu_{\rm true} - \mu_{\rm MLCS} < 0~\rm{ mag},
\end{equation}
which is opposite of the effects noted above.

To assess the impact of these various effects on the inferred
distance, we performed multiple simulations using the SNANA framework
\citep{Kessler09:SNANA}.  We simulated a large sample of \ubv
light curves for SNe~Ia at $z = 0.01$ with $A_{V} = 0$~mag to reduce
the affects of $K$-corrections and extinction.

We fit the light curves in a variety of ways with MLCS2k2.  First, we
fit only the $U$-band data.  As expected
(Section~\ref{ss:singleband}), we find $\Delta M = \mu_{\rm true} -
\mu_{\rm MLCS} > 0$~mag.  Next, we fit the full \ubv and only the $BV$
light curves with $A_{V} = 0$~mag for all SNe.  This isolates any
potential correlations between color-luminosity and color-extinction
relationships.  We find
\begin{equation}
  \mu_{BV} - \mu_{U\!BV} = -0.11~\rm{ mag},
\end{equation}
which is opposite to the effect described in
Section~\ref{ss:singleband}.  Clearly the multi-band data are
important.  If we fix the $U$-band $K$-correction, $K_{UU}$, to zero,
the results hold to within 0.01~mag; therefore, $K$-corrections are
not a dominant factor.

Next, we fit the light curves fixing the light-curve shape parameter,
$\Delta$, for each SN to be the same as the simulated value for that
SN.  In this case, we find
\begin{equation}
  \mu_{BV} - \mu_{U\!BV} = +0.05~\rm{ mag},
\end{equation}
in rough agreement with the results of Section~\ref{ss:singleband}.
We have performed a similar analysis with an SDSS-like simulation
($0.1 < z < 0.3$ and \gri filters) and obtained similar results.

From these findings, we can conclude that including miscalibrated
$U$-band data in the light-curve fits dramatically affects the
relation between light-curve shape and observed color, which in turn
affects the measured distance modulus.  The Keck/SDSS sample has less
UV flux than the low-redshift sample, on average, and therefore has
$\Delta M > 0$~mag.

\section{Subsamples of the Keck/SDSS Sample}\label{s:subsamples}

In Section~\ref{ss:composite}, we created several subsamples of the
entire set of Keck/SDSS SNe to test if various quality cuts affected
the composite spectra.  We determined through this process that we
should exclude SNe where the galaxy subtraction procedure failed.

Since our sample of high-redshift SNe~Ia is relatively large, we are
also able to create subsamples based on physical parameters.  As we
discussed in Section~\ref{ss:comp}, the Nominal Keck/SDSS and
low-redshift composite spectra have very similar average phases and
light-curve shapes, the dominant factors in spectral differences.
Below we will separate the Nominal sample based on other physical
properties to investigate possible physical reasons for the difference
in UV SED between the Keck/SDSS and low-redshift samples.  Some
properties of the subsamples are listed in Table~\ref{t:comp}.

\subsection{Redshift}\label{ss:redshift}

\begin{figure}
\begin{center}
\epsscale{1.05}
\rotatebox{90}{
\plotone{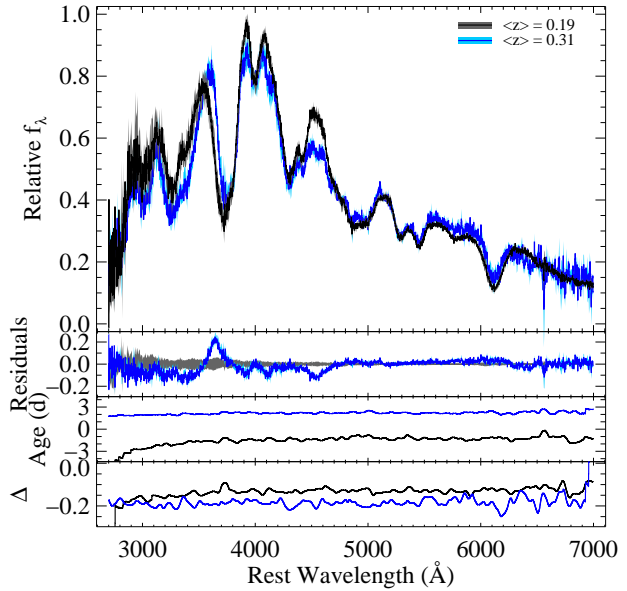}}
\caption{({\it top panel}): Composite spectrum created from Keck/SDSS
SNe with $z < 0.25$ (black curve) and $z > 0.25$ (blue curve).  The
grey and light-blue regions are the 1$\sigma$ boot-strap sampling
errors for the lower-redshift and higher-redshift composite spectra,
respectively.  ({\it second panel}): The grey region is the 1$\sigma$
boot-strap sampling region for the lower-redshift composite spectrum.
The blue curve is the residual of the lower-redshift and
higher-redshift composite spectra.  The light-blue region is the
residual of the lower-redshift composite spectrum and the
higher-redshift 1$\sigma$ boot-strap sampling region.  ({\it third
panel}): The average phase relative to maximum brightness as a
function of wavelength for the lower-redshift (black curve) and
higher-redshift (blue curve) composite spectra.  ({\it bottom panel}):
The average value of $\Delta$ as a function of wavelength for the
lower-redshift (black curve) and higher-redshift (blue curve)
composite spectra.  The UV flux is similar in both
spectra.}\label{f:z_comp}
\end{center}
\end{figure}

Splitting the Nominal sample by redshift (using the mean redshift of
the Nominal sample, 0.25, as the division), we are able to create two
samples: a lower-redshift sample of nine SNe with $\mean{z} = 0.19$,
$\mean{t} = -1.3$~days, and $\mean{\Delta} = -0.12$, and a
higher-redshift sample of eight SNe with $\mean{z} = 0.31$, $\mean{t}
= 2.2$~days, and $\mean{\Delta} = -0.19$.  Composite spectra created
from these samples are shown in Figure~\ref{f:z_comp}.

The composite spectra from the lower-redshift and higher-redshift
subsamples are very similar at most wavelengths.  The major difference
is in the blue wing of the Ca H\&K feature, which extends farther to
shorter wavelengths in the lower-redshift composite spectrum.  As the
velocity of the minimum of the Ca H\&K feature has a range of about
$-$12,000 to $-$22,000~km~s$^{-1}$ at $t = -5$~days and about
$-$11,000 to $-$18,000~km~s$^{-1}$ at $t = 5$~days, with the line
wider at earlier times (e.g., \citealt{Foley11:vgrad}; see also
Section~\ref{ss:sample_selection}), the difference in the two
composite spectra is likely to be an age effect; the lower-redshift
sample has an average phase 2.8~days earlier than the average phase of
the higher-redshift sample.

The other difference between the spectra is at \about 4500~\AA, a peak
between absorption features.  Although it may be attributed to the
different average redshifts, it is more likely that this difference is
also an age effect.

The lower-redshift composite spectrum appears to have slightly more UV
flux than the higher-redshift composite spectrum, but the continua are
still consistent within the boot-strap sampling errors.  Since the
SDSS-II sample should have relatively little Malmquist-associated
biases at $z \approx 0.2$, the lack of a difference in UV SED between
the two composite spectra indicates that these sorts of biases are not
creating the mismatch in SEDs between the low-redshift and Keck/SDSS
samples.

\subsection{Host-Galaxy Star Formation and Mass}\label{ss:mass}

Recently, there have been some indications that SNe~Ia from different
host environments have different peak luminosities {\it after}
correction for light-curve shape and color \citep{Hicken09:de,
Kelly10, Sullivan10, Lampeitl10:host}.  Although these studies have
noticed a correlation between Hubble residuals and both morphology
\citep{Hicken09:de} and mass or star-formation rate \citep{Kelly10,
Sullivan10, Lampeitl10:host}, the cause of the correlation is unclear,
and it is not known whether the correlation is physical or related 
to the light-curve fitting.

As a byproduct of fitting galaxy SEDs to the host-galaxy photometry,
we can obtain a measurement of the mass and star-formation history of
the host galaxies.  The \texttt{kcorrect} routine constructs galaxy
SEDs from several hundred spectral templates.  Using the relative
importance of each template in the galaxy SED, the star-formation
history of the galaxy is reconstructed \citep{Blanton07}.  Once we
determine the luminosity of the galaxy (using the photometry,
redshifts, and a cosmological model), the spectral templates provide a
measurement of the stellar mass of the galaxy.  Although
\texttt{kcorrect} does not directly output an uncertainty for its mass
estimate, the typical uncertainty is 0.1~dex \citep{Blanton07}.

From the Nominal sample, there are four SNe with no host detected in
the SDSS images.  Consequently, we have no information about the
star-formation rate of these galaxies (although they likely have high
specific star-formation rates).  We can, with reasonable confidence,
say that the undetected host galaxies are of relatively low mass.  As
an example, the host of SN~16567 at $z = 0.37$ (the highest redshift
galaxy in the sample) has a host galaxy with $M = 11.8 \times 10^{9}\,
{\rm M}_{\sun}$.  Undetected galaxies at lower redshift ($z = 0.19$,
0.215, 0.29, and 0.33) almost certainly have lower mass.

With the full Nominal sample and the subsample of 13 SNe~Ia having
host-galaxy detections, we are able to respectively split the SNe by
total stellar mass and $B1000$, the percentage of the total star
formation over the previous 1~Gyr compared to the total star formation
over all time. This information is listed in Table~\ref{t:spec}.  The
full sample can easily be divided at $M = 5 \times 10^{9}\, {\rm
M}_{\sun}$ into two groups.  We include SNe~6933 ($z = 0.215$) and
19101 ($z = 0.19$), which do not have detected hosts, in the low-mass
category.  Since SNe~7475 ($z=0.33$) and 19128 ($z = 0.29$) could
conceivably have a mass near our chosen division, we do not include
either SN in either subsample, resulting in a final sample of 15 SNe
from which we examine the SED dependence on host-galaxy mass.  The
subsample of SNe with $B1000$ measurements is divided into two groups,
``active'' and ``passive'' having $B1000 > 46$ and $B1000 < 17$,
respectively.

\begin{figure}
\begin{center}
\epsscale{1.1}
\rotatebox{90}{
\plotone{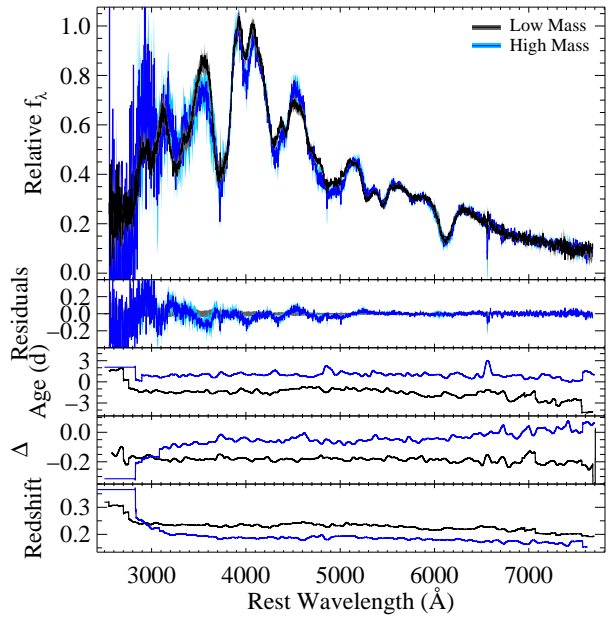}}
\caption{({\it top panel}):  Composite spectra from Keck/SDSS SNe
with host galaxies that have $M < 5 \times 10^{9}\, {\rm M}_{\sun}$
(black curve) and $M > 5 \times 10^{9}\, {\rm M}_{\sun}$ (blue curve).
The grey and light-blue regions are the 1$\sigma$ boot-strap sampling
errors for the low-mass and high-mass composite spectra, respectively.
({\it second panel}): The grey region is the 1$\sigma$ boot-strap
sampling region for the low-mass composite spectrum.  The blue curve
is the residual of the low-mass and high-mass composite spectra.  The
light-blue region is the residual of the low-mass composite spectrum
and the high-mass 1$\sigma$ boot-strap sampling region.  ({\it fourth
panel}): The average phase relative to maximum brightness as a
function of wavelength for the low-mass (black curve) and high-mass
(blue curve) composite spectra.  ({\it fifth panel}): The average
value of $\Delta$ as a function of wavelength for the low-mass (black
curve) and high-mass (blue curve) composite spectra. ({\it bottom
panel}): The average redshift of the low-mass composite spectrum as a
function of wavelength.  The two spectra are relatively similar, but
there is perhaps a slight difference in the UV slope, with the
high-mass composite spectrum having a flatter spectrum from
2900--3500~\AA.}\label{f:ap_comp}
\end{center}
\end{figure}

Not surprisingly, most passive hosts have a large mass and most active
hosts have a small mass.  Because of the larger number of SNe in the
mass segregated sample, we study that one in more detail.  We note
that the same trends appear when using the passive/active subsamples.
We present the composite spectra created from the low-mass and
high-mass subsamples in Figure~\ref{f:ap_comp}.

The low-mass and high-mass composite spectra have average values of
$\mean{z} = 0.23$, $\mean{t} = -1.4$ days, $\mean{\Delta} = -0.18$,
and $\mean{z} = 0.18$, $\mean{t} = 0.9$ days, $\mean{\Delta} = -0.02$,
respectively.  Although the composite spectra from the low-mass and
high-mass subsamples (which are essentially the same as the active and
passive subsamples, respectively) are very similar in the optical,
there are some minor differences in the UV.  In particular, the peak
at \about 3500~\AA\ is higher in the low-mass composite spectrum,
while the peak at \about 2900~\AA\ is lower in the low-mass composite
spectrum.  This results in a relatively flat spectral slope for the
high-mass composite spectrum in the range 2900--3500~\AA.  The
composite spectra do differ by \about 2~days in phase and by \about
0.15 in $\Delta$; however, for the low-redshift sample,
\citet{Foley08:comp} did not see the ``steepness'' of the near-UV
change as a function of mean phase or $\Delta$ (on the other hand,
\citetalias{Foley08:uv} did detect a strong correlation between
$\Delta$ and the slope of the SN's spectrum between 2770 and
2900~\AA).  We therefore believe that it is unlikely that the
discrepancies in the mean phase or $\Delta$ have caused the
differences in these composite spectra.  Unfortunately, the small
sample sizes makes these differences marginally significant, and we
require more SNe to determine whether this trend continues.

Despite the differences in these composite spectra, both have a UV
excess compared to the low-redshift composite spectrum.  It is
therefore unlikely that the UV excess is the result of a biased host
population.

\subsection{Light-Curve Shape}\label{ss:delta}

\begin{figure}
\begin{center}
\epsscale{1.05}
\rotatebox{90}{
\plotone{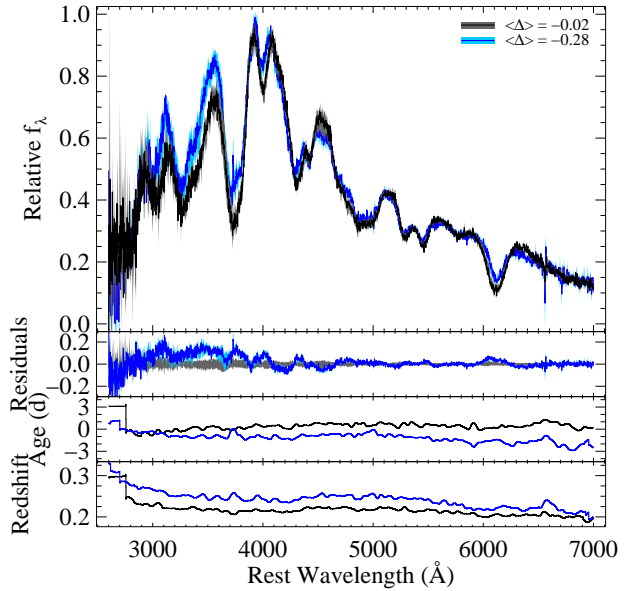}}
\caption{({\it top panel}):  Composite spectra from Keck/SDSS SNe
with $\Delta \ge -0.169$ (black curve) and $\Delta < -0.169$ (blue
curve).  The grey and light-blue regions are the 1$\sigma$ boot-strap
sampling errors for the negative-$\Delta$ and zero-$\Delta$ composite
spectra, respectively.  ({\it second panel}): The grey region is the
1$\sigma$ boot-strap sampling region for the negative-$\Delta$
composite spectrum.  The blue curve is the residual of the
negative-$\Delta$ and zero-$\Delta$ composite spectra.  The light-blue
region is the residual of the negative-$\Delta$ composite spectrum and
the zero-$\Delta$ 1$\sigma$ boot-strap sampling region.  ({\it fourth
panel}): The average phase relative to maximum brightness as a
function of wavelength for the negative-$\Delta$ (black curve) and
zero-$\Delta$ (blue curve) composite spectra.  ({\it bottom panel}):
The average redshift of the negative-$\Delta$ composite spectrum as a
function of wavelength.  Both spectra have similar UV
flux.}\label{f:delta_comp}
\end{center}
\end{figure}

Splitting the Nominal sample by $\Delta$ (using the median value of
the Nominal sample, $\Delta = -0.169$, as the division), we are able
to create two subsamples, one with a smaller average $\Delta$ (the
``Negative $\Delta$'') with eight SNe and average values of $\mean{z}
= 0.24$, $\mean{t} = -1.2$~days, and $\mean{\Delta} = -0.28$, and one
with a larger average $\Delta$ (the ``Zero $\Delta$'') with nine SNe
and average values of $\mean{z} = 0.21$, $\mean{t} = 0.4$~days, and
$\mean{\Delta} = -0.02$.  Composite spectra created from these samples
are shown in Figure~\ref{f:delta_comp}.

The composite spectra from the zero and negative-$\Delta$ subsamples
exhibit many differences at all wavelengths.  The \ion{Si}{2} $\lambda
5972$ feature is weak in the negative-$\Delta$ composite spectrum.
The \ion{Si}{2} $\lambda 6355$ feature is also significantly weaker in
the negative-$\Delta$ composite spectrum compared to the zero-$\Delta$
composite spectrum.  There are also many insignificant differences
redward of \about 4300~\AA.  Clearly the optical SEDs of SNe in these
subsamples are different, but the individual features are similar
enough that the spectra are consistent within the boot-strap sampling
errors.

The spectra continue to deviate at bluer wavelengths.  In addition to
weak \ion{Si}{2} $\lambda 5972$ and \ion{Si}{2} $\lambda 6355$
features, the negative-$\Delta$ composite spectrum has a relatively
weak \ion{Si}{2} $\lambda 4130$ line.  The Ca H\&K feature is also
weaker in the negative-$\Delta$ spectrum.  The blue edge of the Ca
H\&K feature has higher flux in the negative-$\Delta$ spectrum,
indicating that the apparent weakness of Ca H\&K could be caused by
more continuum flux at the position of the feature; however, scaling
the spectra to that peak, the Ca H\&K feature is still weaker in the
negative-$\Delta$ spectrum.  Therefore, we believe that this
difference is inherent to the absorption feature.  These differences
can be described by the $\mathcal{R}($\ion{Ca}{2}$)$ parameter of
\citet{Nugent95}, which is the ratio of the flux at the red edge of
the Ca H\&K feature to that of the flux at the blue edge.  The
Keck/SDSS composite spectra follow the trend of more luminous (or SNe
with smaller $\Delta$) corresponding to smaller
$\mathcal{R}($\ion{Ca}{2}$)$.  These effects have also been seen when
comparing low-redshift composite spectra \citep{Foley08:comp}.

The UV flux is clearly higher in the negative-$\Delta$ spectrum in the
range 3000--3600~\AA.  The absorption minima of the \ion{Fe}{2}
$\lambda$3250 feature is at $-$23,500 and $-$21,300~\kms for the
negative and zero-$\Delta$ spectra, respectively.  This is consistent
with the rough trend that low-$\Delta$ SNe have a higher velocity for
this feature at maximum brightness \citepalias{Foley08:uv}.  The peaks
just blueward and redward of this feature are called $\lambda_{1}$ and
$\lambda_{2}$, respectively, by \citet{Ellis08}.  The
negative-$\Delta$ composite spectrum has a smaller wavelength for the
peak of $\lambda_{2}$ than the zero-$\Delta$ composite spectrum.
\citet{Ellis08} found that there was significant scatter in the
position of these features and that the position was strongly
correlated with phase.

Examining Figure~15 of \citet{Ellis08}, we see that the Keck/SDSS
sample shows similar results when separating the sample into two bins
based on light-curve shape.  Their ``high-stretch'' (corresponding to
smaller $\Delta$) spectrum has stronger UV flux, a smaller
$\mathcal{R}($\ion{Ca}{2}$)$, a higher velocity for the \ion{Fe}{2}
$\lambda 3250$ feature, and a shorter peak wavelength for
$\lambda_{2}$.  All of these characteristics are also seen in the
Keck/SDSS sample.

\section{Investigation of the UV Ratio}\label{s:uvratio}

\begin{figure}
\begin{center}
\epsscale{0.9}
\rotatebox{90}{
\plotone{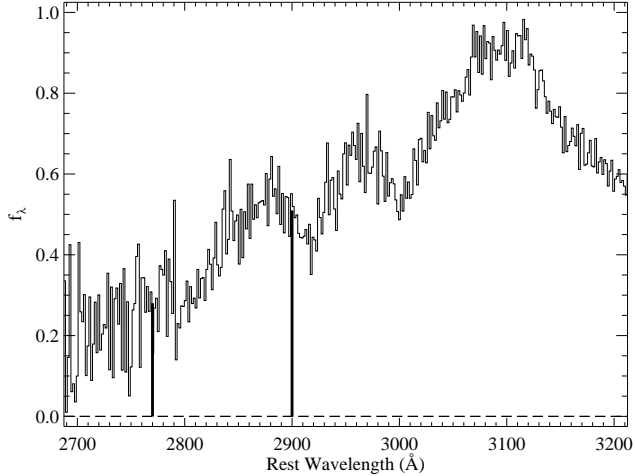}}
\caption{Rest-frame UV spectrum of SN~16618 ($z = 0.193$; $t =
-4.4$~days. Marked are the wavelengths (2770 and 2900~\AA) and
corresponding fluxes used for calculating the UV ratio,
$\mathcal{R}_{UV}$.}\label{f:uvratio}
\end{center}
\end{figure}

The set of Keck/SDSS SN~Ia spectra is comparable to the current
low-redshift UV sample in size and quality \citepalias{Foley08:uv}.
The low-redshift UV sample was mostly obtained with the {\it
International Ultraviolet Explorer} ({\it IUE}) telescope, including
almost all spectra near maximum light (SN~2005cf was observed by {\it
Swift}; \citealt{Bufano09}).  There are 15 low-redshift SN~Ia UV
spectra with $-3 < t < 3$~days from 7 SNe~Ia (SNe~1980N, 1981B, 1986G,
1990N, 1991T, 1992A, 2005cf).  These ``classic'' SNe~Ia all have
precise distance measurements from independent Cepheid observations or
surface brightness fluctuations.  Consequently,
\citetalias{Foley08:uv} were able to determine a relationship between
the ratio of the flux of the SN spectrum at 2770 and 2900~\AA,
$\mathcal{R}_{UV}$ (the ``UV ratio''), and peak luminosity.  An
example of how the UV ratio is calculated is shown in
Figure~\ref{f:uvratio}.

The UV ratio is an excellent luminosity indicator for these six
low-redshift SNe~Ia.  Using a linear relationship, we see that the UV
ratio yields a scatter in the peak absolute magnitudes of the SNe of
0.21~mag, comparable to the scatter found when using the luminosity
vs.\ light-curve shape relationship ($\sigma \approx 0.17$ for this
sample; \citetalias{Foley08:uv}).  Furthermore, combining the UV ratio
and the luminosity vs.\ light-curve shape relationship,
\citetalias{Foley08:uv} were able to reduce the scatter to 0.09~mag.

\begin{figure}
\begin{center}
\epsscale{0.9}
\rotatebox{90}{
\plotone{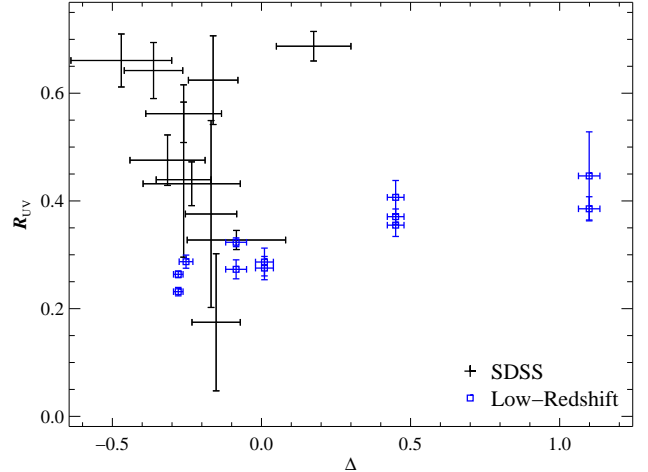}}
\caption{Relationship between the UV ratio ($\mathcal{R}_{UV}$) and
$\Delta$.  The black points represent the SNe in our Keck/SDSS
sample.  The blue squares are the low-redshift sample from
\citetalias{Foley08:uv}.  The Keck/SDSS SNe are obviously offset
to higher values of $\mathcal{R}_{UV}$, the result of the UV excess.}
\label{f:delta_ratio}
\end{center}
\end{figure}

For a subsample of our Keck/SDSS SNe, we possess accurately
fluxed-calibrated UV spectra (based on our galaxy-subtraction
techniques) that have reasonable S/N at 2770~\AA\ to measure
$\mathcal{R}_{UV}$.  These SNe are all SNe from the Nominal sample
observed in 2006 and 2007, supplemented by those with $z \ge 0.215$
from 2005.  In Figure~\ref{f:delta_ratio}, we present the measured UV
ratios from our sample as a function of $\Delta$ and compare to the
low-redshift values.  The SDSS-II light curves were fit with a
slightly different version of MLCS2k2 than that of the low-redshift
sample (see Section~\ref{ss:spec}).  Fitting a subset of low-redshift
SNe with both versions of MLCS2k2, we see an offset of 0.01 and a
root-mean-square scatter of 0.02 in $\Delta$ between the two outputs.
The larger uncertainty in $\Delta$ for the Keck/SDSS SNe is the result
of slightly lower quality (both S/N and temporal sampling) light
curves.  The larger uncertainty in $\mathcal{R}_{UV}$ for the
Keck/SDSS SNe is due to the combination of slightly lower-S/N spectra
and larger values of $\mathcal{R}_{UV}$ (for spectra with a fixed S/N,
one naturally expects a larger uncertainty for larger values of
$\mathcal{R}_{UV}$).

Examining Figure~\ref{f:delta_ratio}, we immediately notice that for a
given $\Delta$, the Keck/SDSS sample has a higher value of
$\mathcal{R}_{UV}$.  This is expected in view of the UV excess in the
SDSS-II sample noted above; the UV excess is more prominent at shorter
wavelengths.  Unfortunately, the small range of $\Delta$ for the
Keck/SDSS sample does not allow us to determine whether the trend seen
at low redshift (i.e., $\mathcal{R}_{UV}$ increases with $\Delta$)
also exists at high redshift.

As a test of using $\mathcal{R}_{UV}$ as a luminosity indicator, we
wish to determine $M_{V}$ for each Keck/SDSS SN~Ia and compare that to
$\mathcal{R}_{UV}$.  To find $M_{V}$, we first measure the peak
apparent $r$ or $i$ magnitude (depending on the redshift of the SN)
using MLCS template light curves fit to the photometry in that single
band.  Then, using the concordance cosmology ($H_{0} =
70$~km~s$^{-1}$~Mpc$^{-1}$, $\Omega_{m} = 0.3$, $\Omega_{\Lambda} =
0.7$, and $w = -1$), we can determine the distance modulus ($\mu$) of
each SN.  From our MLCS fits and \citet{Schlegel98}, we are able to
determine the host-galaxy and Milky Way reddenings, $A_{V, {\rm
host}}$ and $A_{X,{\rm MW}}$, respectively.  Finally, from our MLCS
fits, we are also able to derive a $K$-correction, K$_{XV}$, where
$XV$ denotes a correction from observed-band $X$ to rest-frame
$V$. Then, using the equation
\begin{equation}
  M_{V} = X_{\rm max} - \mu - A_{V, {\rm host}} - A_{V,{\rm MW}} - 
{\rm K}_{XV},
\end{equation}
where $X$ denotes either the $r$ or $i$ band, we are able to determine
the absolute visual magnitude for each SN. Note that there are
assumptions made regarding the $K$-corrections which may not be
appropriate for our sample given the differences between the spectra
of low-redshift and Keck/SDSS SNe seen above.  However, the rest-frame
$V$ band is relatively free of these differences.  Another assumption
is that the MLCS-measured value for $A_{V}$ applies to our sample
given the observed color differences between the high and low-redshift
samples.  However, uncertainties in $A_{V}$ and K$_{rV}$ are typically
\about 0.05 and 0.02~mag, respectively, which are generally smaller
than the errors associated with the determination of $r_{\rm max}$ and
$\mu$.

\begin{figure}
\begin{center}
\epsscale{0.9}
\rotatebox{90}{
\plotone{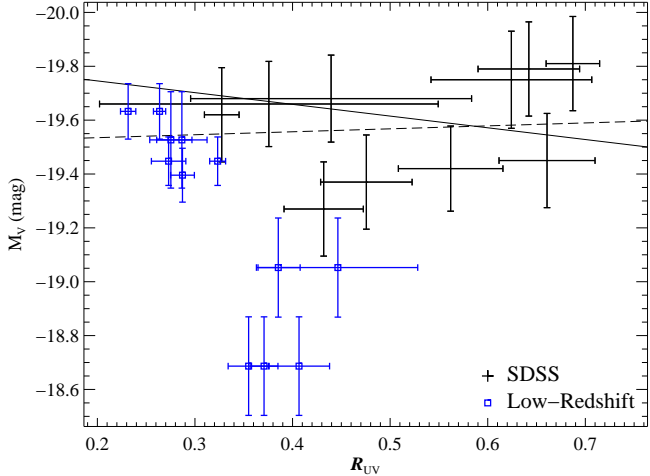}}
\caption{Relationship between the UV ratio ($\mathcal{R}_{UV}$) and
$M_{V}$ at peak brightness.  The black points represent the SNe in our
Keck/SDSS sample.  The blue squares are the low-redshift sample from
\citetalias{Foley08:uv}.  There are multiple measurements for some
low-redshift SNe (from different spectra over $-3 < t < 3$~days).
Each low-redshift SN has a unique value of $M_{V}$.  The solid black
line is the relationship given in Equation~\ref{e:ruv}.  The dashed
black line is the relationship given in Equation~\ref{e:ruv2},
including the low-redshift data with $M_{V} < -19.0$~mag,
corresponding to smaller $\Delta$.  The Keck/SDSS SNe follow a
different trend than the low-redshift SNe that cannot be attributed to
only the UV excess, which would shift the SNe to higher values of
$\mathcal{R}_{UV}$ (and potentially change the slope slightly).
However, it is possible that all SNe with $M_{V} < -19.0$~mag follow a
simple relationship.}
\label{f:mv_ratio}
\end{center}
\end{figure}

We compare the values of $\mathcal{R}_{UV}$ and $M_{V}$ for the
Keck/SDSS and low-redshift SNe in Figure~\ref{f:mv_ratio}.  As found
by \citetalias{Foley08:uv}, the low-redshift SNe follow the
relationship
\begin{equation}\label{e:ruv0}
  M_{V} (t = 0) = -19.366 + 5.686 (\mathcal{R}_{UV} - 0.3)~{\rm mag},
\end{equation}
where we have adjusted $M_{V}$ for the SNe to match $H_{0} =
70$~km~s$^{-1}$~Mpc$^{-1}$.  It is obvious that the trend seen in the
low-redshift sample is not present in the Keck/SDSS sample.  One might
expect a shift of the trend to higher $\mathcal{R}_{UV}$, but a
similar slope, if the UV excess is equal for all Keck/SDSS SNe.
However, fitting all Keck/SDSS SNe with spectra that reach rest-frame
2750~\AA, we find
\begin{equation}\label{e:ruv}
  M_{V} (t = 0) = -19.702 + 0.435 (\mathcal{R}_{UV} - 0.3)~{\rm mag},
\end{equation}
which is much shallower than the low-redshift trend.  The fit
corresponding to Equation~\ref{e:ruv} has $\chi^{2} /\dof = 15.9/11$.

The low-redshift sample contains only six SNe, while the Keck/SDSS
sample consists of SNe from a small range of $M_{V}$ (and $\Delta$).
Four of the low-redshift SNe appear to be consistent with the
relationship derived from the Keck/SDSS SNe, while two (those with the
highest value of $\Delta$, and corresponding to the five low-redshift
points with $M_{V} > -19.1$~mag in Figure~\ref{f:mv_ratio}) are
inconsistent.  If we examine all SNe (at both low and high redshift)
with $M_{V} < -19.1$~mag, we determine
\begin{equation}\label{e:ruv2}
  M_{V} (t = 0) = -19.546 - 0.110 (\mathcal{R}_{UV} - 0.3)~{\rm mag},
\end{equation}
with $\chi^{2} /\dof = 25.9/18$.  The values for the linear fits for
Equations~\ref{e:ruv0}--\ref{e:ruv2} and corresponding uncertainties
are presented in Table~\ref{t:ruv}.

\begin{deluxetable}{lrr}
\tablewidth{0pt}
\tablecaption{$M_{V} - \mathcal{R}_{UV}$ Linear Fits\label{t:ruv}}
\tablehead{
\multicolumn{3}{c}{$M_{V} (t = 0) = A + B (\mathcal{R}_{UV} - 0.3)~{\rm mag}$} \\
\colhead{Equation} &
\colhead{$A$} &
\colhead{$B$}}

\startdata

\ref{e:ruv0} & $-19.366$ (0.298) & \phs5.686 (0.990) \\
\ref{e:ruv}  & $-19.702$ (0.167) & \phs0.435 (0.331) \\
\ref{e:ruv2} & $-19.546$ (0.111) &  $-0.110$ (0.287)

\enddata

\end{deluxetable}

Both samples must be expanded to determine whether (a) these trends
are real within their own populations, (b) the low and high-redshift
samples have truly different relationships, and (c) subsamples of SNe
form groups in the $\mathcal{R}_{UV}$--$M_{V}$ plane.


\section{Discussion}\label{s:disc}

To summarize our results, we find (Section~\ref{s:comp}) that the
intermediate-redshift Keck/SDSS sample has significantly more flux
($>20\%$) in the near-UV ($\lambda < 3400$~\AA) relative to a
low-redshift sample.  This mismatch has significant implications for
measuring distances with SNe~Ia (Section~\ref{s:cosmo}).  In
Section~\ref{s:subsamples}, we examine subsamples of the Keck/SDSS
sample to determine if there were any obvious correlations between
spectral features and redshift, host properties, and light-curve
shape, respectively.  We find that the UV SED of the Keck/SDSS sample
does not depend on redshift, and there are no significant differences
that correlate with host-galaxy mass or light-curve shape.

We now investigate possible systematic errors which could create a UV
mismatch.  We also discuss this result in the context of SN evolution.

\subsection{Possible Systematic Errors}

There are many possible systematic errors associated with the spectral
shape of the SNe~Ia in our study.  They can be separated into three
categories: sample selection, spectral reductions, and post-processing
techniques.  We address each of these in turn.

\subsubsection{Sample Selection}\label{ss:sample_selection}

The above results rely on the assumption that the Keck/SDSS and
low-redshift samples were randomly selected from their respective
populations. Here we discuss a possible source of systematic
uncertainty from selection bias.  If the Keck/SDSS sample presented
here is not representative of high-redshift SNe~Ia, then any
conclusions drawn from this sample are possibly incorrect.  Similarly,
if the sample used to create the low-redshift composite spectrum is
biased relative to the sample of SNe used in a cosmological analysis,
the same caveats apply.

\begin{figure}
\begin{center}
\epsscale{0.65}
\rotatebox{90}{
\plotone{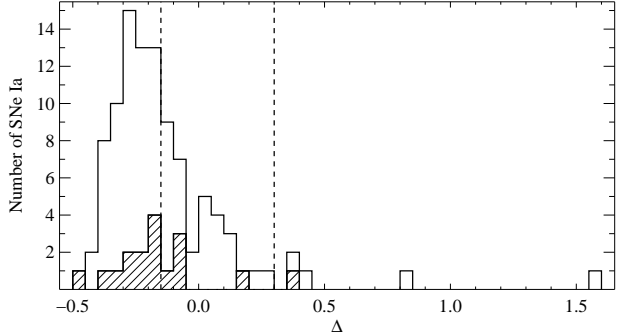}}
\caption{Histogram of the $\Delta$ distribution of the SDSS-II SNe~Ia
from the first SDSS-II cosmological analysis and the subsample
presented in this paper (dashed histogram).  The dashed lines mark the
regions of overluminous ($\Delta < -0.15$), normal ($-0.15 < \Delta <
0.3$), and underluminous ($\Delta > 0.3$) SNe, as defined by
\citet{Jha07} and shown in Figure~\ref{f:hist}.}\label{f:sdss_delta}
\end{center}
\end{figure}

If the Keck/SDSS sample is not a random subset of the larger
spectroscopic SDSS-II sample, then we would expect there to be some
additional observational evidence besides a UV excess.  In
Figure~\ref{f:sdss_delta}, we compare the light-curve shape parameters
of our subsample to the 99 SNe~Ia used in the first SDSS-II
cosmological analysis \citepalias{Kessler09:cosmo}.  Performing a
Kolmogoroff-Smirnov (K-S) test, we find that the two samples are not
significantly different, with a $p$-value of 0.348.  Therefore, the
Keck/SDSS sample appears to be representative (in terms of light-curve
shape) of the overall SDSS-II spectroscopic sample.

\begin{figure}
\begin{center}
\epsscale{1.2}
\rotatebox{90}{
\plotone{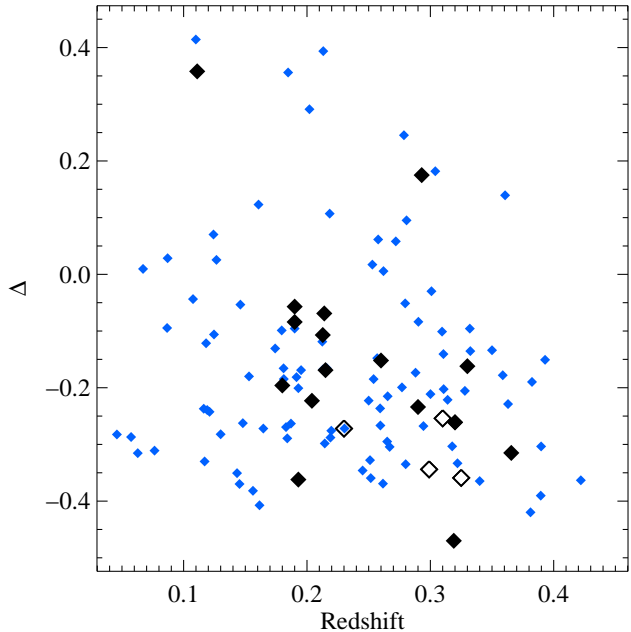}}
\caption{The $\Delta$ distribution of Keck/SDSS SNe~Ia presented in
this paper as a function of redshift.  The filled black diamonds are
the Keck/SDSS SNe in the Nominal sample.  The open black diamonds are
the Keck/SDSS SNe rejected from the Nominal sample.  The smaller blue
diamonds are the full SDSS-II first-year sample.  A K-S test shows
that the Keck/SDSS and full SDSS-II samples are
consistent.}\label{f:zdelta}
\end{center}
\end{figure}

In Figure~\ref{f:zdelta}, we show the measured value of $\Delta$ for
each SN as a function of redshift.  There is no real indication of
Malmquist bias.  A best-fit line to the $\Delta$ distribution of the
Nominal sample with redshift is 1.1$\sigma$ from zero slope.  This,
combined with the changing demographics of the SN population and the
effect on average light-curve shape \citep{Howell07}, suggests that
there is no significant Malmquist bias.

The full, photometric SDSS-II sample is certainly Malmquist biased,
but the SDSS-II sample should be nearly unbiased to $z \approx 0.3$
for discoveries \citepalias[as opposed to spectroscopic completeness,
which is limited to $z \approx 0.15$;][]{Kessler09:cosmo}; the
Keck/SDSS sample may be similarly Malmquist biased, although we do not
strongly detect it.

Besides a simple Malmquist bias where we are selecting more-luminous
SNe, we may be selecting bluer SNe than the typical high-redshift
SN~Ia included in cosmological analyses.  This is particularly
problematic for $z > 1$ SNe~Ia where searches are usually performed in
the rest-frame near-UV (but photometrically followed in redder bands).
For the SDSS-II sample, the search is performed in \gri\!\!\!,
avoiding the near-UV, which should reduce such a bias.  The difference
between $g-r$ color of the Keck/SDSS and low-redshift composite
spectra is $\Delta (g-r) = 0.1$~mag at $z = 0.35$, and less for lower
redshifts.  Therefore, it is highly unlikely that such a bias is
directly affecting our sample.

Nevertheless, the Keck/SDSS sample has an excess of rest-frame UV flux
relative to the low-redshift sample, which in turn has an excess of
rest-frame UV flux relative to the full SDSS-II sample
\citepalias{Kessler09:cosmo}.  Examining the $g$-band MLCS2k2
light-curve fits for the \citetalias{Kessler09:cosmo} sample, we can
assign a value, which we call $p_{\rm UV}$, of $-1$, 0, or 1 to light
curves that are below, consistent with, and above the MLCS2k2 $g$-band
model, respectively.  Our results are shown in Table~\ref{t:gband}.
With increasing redshift, corresponding to bluer wavelengths, the
$g$-band data of the SDSS-II sample go from being above the model to
significantly below the model.  The 12 SNe from the Keck/SDSS sample
that are also in the \citetalias{Kessler09:cosmo} sample, on the other
hand, have $\langle p_{\rm UV} \rangle = 0.23$ (indicating that the
sample tends to be above the model), with $\langle z \rangle = 0.250$.
For the 0.2--0.3 redshift bin, the \citetalias{Kessler09:cosmo} has
$\langle z \rangle = 0.250$ (the same as the Keck/SDSS sample within
the \citetalias{Kessler09:cosmo} sample), but $\langle p_{\rm UV}
\rangle = -0.31$.  Although there is no obvious color selection bias
that causes this effect, the differences appear to be real.  It is
therefore important to examine other differences between the various
samples.

\begin{deluxetable}{lcccc}
\tablewidth{0pt}
\tablecaption{SDSS-II $g$-Band Data/Model Properties\label{t:gband}}

\startdata

\hline
\hline
Redshift Range                             & 0--0.1 & 0.1--0.2 & 0.2--0.3 & $> 0.3$ \\
Number of SNe                              & 10     & 34       & 35       & 23 \\
$\langle z \rangle$                        & 0.072  & 0.153    & 0.250    & 0.349 \\
$\langle p_{\rm UV} \rangle$               & 0.20   & 0.15     & $-0.31$  & $-0.38$ \\
$\langle \lambda \rangle$\tablenotemark{a} & 4760 & 4420 & 4080 & 3780

\enddata

\tablenotetext{a}{Approximate rest-frame wavelength of the center of
the $g$ band at the average redshift for the redshift bin.}

\end{deluxetable}

The low-redshift sample has few UV spectra, and it is possible that
these SNe are not representative of low-redshift SNe~Ia.  The UV
portion of the maximum-light composite spectrum consists of 15 spectra
from 7 SNe (\citetalias{Foley08:uv}; SNe~1980N, 1981B, 1986G, 1990N,
1991T, 1992A, and 2005cf), all observed with {\it IUE} or {\it Swift}.
The low-redshift composite spectrum begins to deviate from the
Keck/SDSS composite spectrum at \about 4000~\AA, which is easily
observed from the ground (and not included in any {\it IUE} spectra),
is not heavily affected by the atmosphere, and is sampled by many
spectra.  We therefore do not think that the small sample of
low-redshift UV spectra is biasing our results.  Although a larger
number of low-redshift UV spectra of SNe~Ia is desperately needed (and
is being addressed by our {\it Swift} program\footnote{PI Filippenko,
Programs 04047, 5080130; PI Foley, Program 6090689.} and complementary
{\it HST} programs\footnote{PI Ellis, Programs 11721 and 12298; PI
Foley, Program 12592.}, a biased low-redshift set of SNe~Ia having
available UV spectra does not appear to be the cause of the UV
difference.

Alternatively, the low-redshift and Keck/SDSS samples might have intrinsic
differences that may cause discrepancies in UV spectra but not show
differences in light-curve shape.  To properly compare the samples, we
must consider the intrinsic differences of the surveys that discover
SNe at low redshift compared to that of SDSS-II.  The predominant
method of finding SNe at low redshift is through targeted surveys,
where a sample of galaxies is searched continually.  In particular,
this is the method used by the Lick Observatory Supernova Search
\citep[LOSS;][]{Filippenko01, Filippenko05b}, from which a large
fraction of the low-redshift SNe~Ia come.  SDSS-II, on the other hand,
is an untargeted survey, observing the same part of the sky
continually, but not targeting particular galaxies.  This means that
the sample demographics are different.  Although most of the stars in
an untargeted survey are in galaxies similar to those searched by
targeted surveys, so they are not dramatically different, there may
still be a significant difference in the host-galaxy population.
Specifically, the host galaxies of the Keck/SDSS sample have a median
mass of (2.6--3.6) $\times 10^{9}\, {\rm M}_{\sun}$ (depending on how
one treats the host-galaxy nondetections), while the average masses
for E, Sb, and Irr galaxies from LOSS, from which a significant number
of the low-redshift sample come, are 16.0, 6.4, and $2.4 \times
10^{10}\, {\rm M}_{\sun}$ \citep{Leaman10}.  Therefore, the average
SN~Ia from the Keck/SDSS sample is hosted in a galaxy that is probably
\about 0.1 times the mass of the average {\it irregular} galaxy and
\about 0.02 times the mass of the average elliptical galaxy in the
low-redshift sample.  Given the mild differences found in the spectra
of SNe hosted in low and high-mass hosts (see Section~\ref{ss:mass})
and the correlation between host-galaxy mass and Hubble residuals
\citep{Kelly10, Lampeitl10:host, Sullivan10}, we consider variations
imprinted on SNe~Ia by their environments as a possible cause of the
difference.  However, the Keck/SDSS sample is a subset of the SDSS-II
sample, and it is unclear how it could have significantly different
average host properties.  Future untargeted, low-redshift SN samples
from surveys such as the Palomar Transient Factory (PTF), La
Silla-QUEST, SkyMapper, or Pan-STARRS should provide SNe with similar
host-galaxy properties as the high-redshift SNe.  (However, the
low-redshift sample of \citealt{Cooke11} primarily came from PTF and
showed similar UV flux as our low-redshift sample.)

A second difference between the searches is that those at low redshift
tend to be done in a single band (often unfiltered, which for LOSS is
similar to the $R$ band), while the SDSS-II search is performed in
\gri\!\!.  The SDSS-II search requires a detection in two of the
filters.  At $z \approx 0.25$, these are typically $r$ and $i$.  It is
therefore unlikely that the Keck/SDSS sample is biased to select
significantly bluer SNe at discovery than for the low-redshift sample.
In a typical exposure, the SDSS-II SN survey achieved a 50\% detection
limit of $g = 22.7$~mag, which is \about0.5~mag deeper than the
typical $z = 0.35$ SN~Ia at peak.  As mentioned above, the difference
between the $g-r$ color for the Keck/SDSS and low-redshift composite
magnitudes is $\Delta (g-r) = 0.10$~mag at $z = 0.35$ (and $\Delta
(g-r) < 0.2$~mag until $z \approx 0.55$), so we should not be
significantly affected by this potential bias.  Additionally, the
rest-frame $g-r$ colors of the Nominal and low-redshift composite
spectra is only 0.069~mag.

Examining Figure~\ref{f:hist}, we see that the Keck/SDSS sample is
predominantly composed of SNe with broad light curves (i.e., low
$\Delta$).  Although the full low-redshift and Keck/SDSS samples have
different demographics, it is important to point out that (a) this is
a natural consequence of the evolving demographics of SNe~Ia with
redshift \citep[e.g.,][]{Howell07}, and (b) the low-redshift sample we
have used to generate the comparison low-redshift composite spectrum
has very similar properties to that of the Keck/SDSS sample (as
demonstrated by the comparison of phase and $\Delta$ in
Figure~\ref{f:comp}). Consequently, differences in the spectra as a
result of different photometric and phase properties should be largely
mitigated and not biasing our findings.  However, note that we draw
some conclusions in later sections based on the full low-redshift
photometric sample which may not show a similar trend.

\subsubsection{Spectral Reductions}

Another set of possible systematic errors is associated with the
observations and reduction process.  The two major sources of
systematic error would be the systematic ``reddening'' of the spectra
as the result of the spectroscopic slit not being aligned along the
parallactic angle \citep{Filippenko82} and incorrect flux calibration.

All of the Keck observations were taken within $5^\circ$ of the
parallactic angle to preserve the intrinsic SED of the SN spectrum.
Furthermore, the observations taken in 2007 were with the ADC in
place, reducing the effective atmospheric dispersion to a negligible
amount for all but the highest airmasses. Moreover, if significant
atmospheric dispersion were present at high airmass, the blue flux
should decrease more than the red, given that the SNe were centered in
the slit while using the red-sensitive LRIS guide camera --- yet this
is the opposite of the observed difference between the low and
high-redshift samples. (Note that the standard stars were observed in
the same manner as the SNe.) In addition, the low-redshift UV spectra
were obtained from space, where atmospheric dispersion does not affect
the observations.

The second possibility is that the flux calibration of the SN spectra
was incorrect, causing an overestimate of the bluest fluxes.  However,
the spectra were obtained over three seasons on several nights.
Therefore, any flux-calibration error must have occurred for these
temporally separated observations.  The Keck/SDSS SN spectra were also
reduced in exactly the same manner as that of the optical spectra from
the low-redshift sample (and similar in methodology for the UV
spectra), with all Keck/SDSS SN spectra and a significant number of
low-redshift SN spectra reduced by the same person (R.J.F.).  If there
is an inherent flux-calibration error for all spectra, this would have
been observed in the low-redshift sample as well.  A second set of
spectral reductions was performed by another person (Ryan Chornock)
with similar results.  Additionally, the spectra obtained in 2005 were
observed with polarimetry optics, producing two spectra for each
object that were reduced with completely separate calibrations, yet
they always agreed within the uncertainties.  A final argument against
a flux-calibration error is that these errors would be present in the
{\it observer} frame, not in the rest frame where the difference is
apparent.  Since our spectra span a range in redshift, the errors
should be present at varying wavelengths and not correspond to the
same rest-frame wavelengths.  However, when separating our sample by
redshift, we see no difference (see Section~\ref{ss:redshift}).
Because of our meticulous flux calibration with multiple decades of
testing, flux-calibration errors seem unlikely.

\subsubsection{Post-Processing Techniques}

A final candidate for systematic errors resides in the post-processing
techniques performed on the spectra.  The possibilities include
incorrect extinction corrections, galaxy-contamination corrections,
and creation of the composite spectra.

There are two separate extinction corrections applied to the spectra.
The first is a Milky Way extinction correction, using the dust maps of
\citet{Schlegel98}.  The extinction measurements from these maps are
typically quite good.  Furthermore, the Keck/SDSS SNe tend to come
from low-extinction regions ($E(B-V) \lesssim 0.05$~mag).  It is
therefore unlikely that the Milky Way extinction corrections are
sufficiently erroneous to significantly influence the measured flux.

The second extinction correction is to account for the line-of-sight
extinction due to host-galaxy dust.  The extinction has been estimated
during the MLCS fitting.  For our Keck/SDSS sample, the estimates are
typically low (median value for the Nominal sample is $A_{V} =
0.04$~mag and the maximum value is $A_{V} = 0.43$~mag).  However, the
reddening is determined by comparing the high-redshift SN light curves
to template light curves derived from a low-redshift training
set. Consequently, if there is a difference between the low and
high-redshift samples, this estimate may be incorrect.  On the other
hand, since the extinction estimates are low, we are making small
corrections to the spectra, and it is difficult to explain how these
small corrections could create the large UV excess in the Keck/SDSS
composite spectrum.  Finally, \citet{Foley08:comp} showed that
composite spectra formed from only low-extinction, uncorrected spectra
and composite spectra formed from both low and high-extinction,
extinction-corrected spectra are consistent with each other.

\begin{figure}
\begin{center}
\epsscale{0.9}
\rotatebox{90}{
\plotone{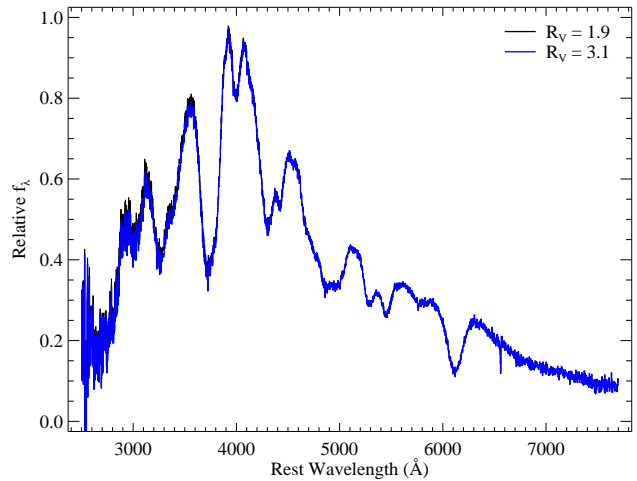}}
\end{center}
\caption{Composite spectra from Keck/SDSS SNe in the Nominal
sample with $R_{V} = 1.9$ (black curve) and $R_{V} = 3.1$ (blue curve)
for all light-curve fits and extinction corrections.  The spectra are
nearly indistinguishable, indicating that the choice of $R_{V}$ does
not affect the UV excess.}\label{f:rv_comp}
\end{figure}

The light curves of the Keck/SDSS sample were fit with $R_{V} = 1.9$,
the best-fit value found for the full SDSS-II sample\footnote{This
value of $R_{V}$ was determined by comparing the SDSS-II photometry to
simulations; see Section~7.2 of \citet{Kessler09:cosmo}.}.  However,
this value of $R_{V}$ is low compared to that seen in the Milky Way
\citep{Schlegel98}, which may be partially the result of mixing SNe~Ia
with different intrinsic colors \citep{Foley11:vel}.  We have
performed a separate set of light-curve fits with $R_{V} = 3.1$.
Since the value of $A_{V, {\rm host}}$ is low for the Keck/SDSS SNe,
changing the value of $R_{V}$ did not dramatically alter the values of
$A_{V, {\rm host}}$.  In Figure~\ref{f:rv_comp}, we show the Keck/SDSS
composite spectrum presented in Figure~\ref{f:comp} (created from the
Nominal sample with $R_{V} = 1.9$ for all light-curve fits and our
extinction corrections) and a composite spectrum created from the
Nominal sample with $R_{V} = 3.1$ for all light-curve fits and
extinction corrections.  These two spectra are created from the same
sample of SNe, and the only difference is in the value of $R_{V}$ and
derived $A_{V, {\rm host}}$.  Although the different value for $R_{V}$
does change the SED of the composite spectrum slightly, it is not a
sufficiently large difference to compensate for the UV excess.

As shown in Figure~\ref{f:galsub}, the galaxy-contamination correction
appears to be quite good for the observer-frame \gri bands.  These
measurements do not include the observer-frame near-UV region, where
the rest-frame UV is observed.  However, the galaxy spectra are
determined from \ugriz photometry, which does include the
observer-frame near-UV region, and other than at $0.27 < z < 0.33$,
the reconstruction of galaxy SEDs is very good \citep{Blanton03}.
Finally, as seen in Figure~\ref{f:comp_diff}, the composite spectra
created from the full Keck/SDSS sample presented in this paper and
from the sample of SNe with little or no galaxy contamination are
consistent.  It is thus unlikely that incorrect galaxy contamination
causes the UV excess.

The final possible source of post-processing systematic error is
related to creating the composite spectra.  Since the low-redshift and
Keck/SDSS composite spectra were created in exactly the same way,
systematic differences should be minimal.  The one major possible
systematic error associated with creating the composite spectra comes
from splicing some low-redshift spectra together.  Unlike with the
Keck/SDSS spectra, which all contain a large overlapping rest-frame
wavelength range, some low-redshift spectra are completely disjoint.
Accordingly, we splice the spectra together using an iterative
process, creating composite spectra, rescaling spectra, and recreating
composite spectra until all spectra are added.  This is especially
important for those particular UV spectra whose longest wavelengths
are \about 3200~\AA.

\begin{figure}
\begin{center}
\epsscale{0.9}
\rotatebox{90}{
\plotone{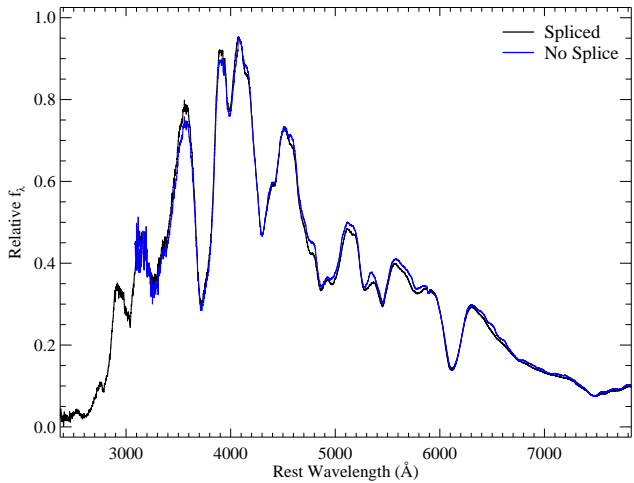}}
\caption{Low-redshift composite spectra with (black curve) and without
(blue curve) splicing spectra (as described in the text).  The small
differences between the spectra are mainly the result of slightly
different samples and the iterative scaling of each spectrum allowed
in the splicing method.  The spectra are extremely similar for the
overlapping wavelengths, indicating that splicing the UV spectra to
the optical spectra does not affect the flux at the shortest optical
wavelengths.}\label{f:splice}
\end{center}
\end{figure}

For most composite spectra, the overlap between the UV and
ground-based optical spectra is \about 100~\AA.  Consequently, there
may be an offset between the UV and optical portions of the composite
spectrum at \about 3200~\AA\ (the number of spectra for the composite
spectra as a function of wavelength can be seen in
Figure~\ref{f:comp}).  We have created a composite spectrum without
splicing any spectra together (using only the SNe that overlap over
the same 1000~\AA\ region), and compare in Figure~\ref{f:splice} that
spectrum to our composite spectrum consisting of all spectra.  From
this figure we see that the method of splicing spectra together does
not significantly change the composite spectrum over the wavelength
region covered by both spectra.  Figure~\ref{f:splice} also shows that
the UV spectra appear to be spliced properly, creating the correct
SED.  However, we cannot rule out the possibility of errors at the
ends of our spectra combining to incorrectly set the scaling of the UV
spectra.  Regardless of these difficulties, there are differences
between the Keck/SDSS and low-redshift composite spectra at
wavelengths longer than 3200~\AA, indicating that there are probably
some real differences between these samples.  With additional
UV-optical spectra obtained at {\it exactly} the same epoch, we may be
able to address this further.

\subsection{Supernova Evolution}

No apparent systematic error is creating the UV excess for the
Keck/SDSS SN~Ia sample.  We now investigate the possible physical
reasons for a UV excess in the high-redshift SNe~Ia compared to
low-redshift SNe~Ia.  Changes in SNe~Ia with redshift are typically
called ``evolution,'' but it is possible that the sample demographics
are simply shifting with redshift (see \citealt{Foley08:comp} and
\citealt{Leibundgut01} for discussions).  We will not distinguish
between these scenarios here, but we point out that if the cause
of the difference is related to host-galaxy mass (or star-formation
rate, etc.), which is compounded by the differences between targeted
(at low redshift) and untargeted (at high redshift) SN searches, this
would not be considered evolution.

\citetalias{Kessler09:cosmo} found that including and excluding the
observed $u$ band for nine low-redshift ($0.04 < z < 0.09$) SDSS-II
SNe resulted in the same distance modulus offset as found for the rest
of the SDSS-II sample when including and excluding the rest-frame $U$
band.  The non-SDSS-II low-redshift sample ($z \approx 0.03$ for most
SNe) did not show this offset; if evolution is causing a difference in
the SEDs, the evolution must occur quickly and recently.

There are several models for SNe~Ia evolving with redshift as the
result of changing progenitor metallicity \citep[e.g.,][]{Hoflich98,
Lentz00, Sauer08}.  A major conclusion of these studies is that the
optical spectra will have relatively small differences, while the UV
will show a significant difference.  Unfortunately, there is no
consensus as to whether the UV flux will increase or decrease with
increasing metallicity.

The models of \citet{Hoflich98} indicate that UV excess is expected
with higher progenitor metallicity.  \citet{Lentz00}, however, showed
model spectra where UV excess is the result of lower progenitor
metallicity.  The differences are the result of differing density
structures \citep{Lentz00, Dominguez01}.  It is beyond the scope of
this study to determine which model is more accurate.  Although other
factors such as the progenitor C/O ratio may influence the UV flux,
the difference in UV spectra appears to be broadly consistent with
models of differing metallicity.

The slight apparent difference between SN~Ia UV spectra from SNe with
active/low-mass and passive/high-mass host galaxies indicates that
SNe~Ia with different progenitors or delay times may result in
different explosions despite having similar light-curve shapes.  It is
necessary to obtain further observations to increase the sample size
and see if this trend persists.

There is a clear difference in the relationship between
$\mathcal{R}_{UV}$ and $M_{V}$ for the low-redshift and Keck/SDSS
samples.  It is possible that this is a sign of evolution.  However,
neither the low nor high redshifts have been adequately sampled in the
full $M_{V}$--$\Delta$ plane.  We must therefore expand these samples
to determine if these correlations persist.  However, at this time the
relationship between $\mathcal{R}_{UV}$ and $M_{V}$ found for the
low-redshift sample by \citetalias{Foley08:uv} clearly should {\it
not} be used for determining the peak luminosity of a high-redshift
SN~Ia.

When comparing the Keck/SDSS and low-redshift samples, we have
controlled for most observables.  In particular, the composite spectra
shown in Figure~\ref{f:comp} have the same average phase and
light-curve shape.  The one obvious observable which is vastly
different between the samples is host-galaxy mass.  As we discussed
above, the average host for an SN in the low-redshift sample is
probably 10--50 times more massive than that of an SN in the Keck/SDSS
sample.  Given the similarity in other parameters, we wonder if such a
difference is the dominant factor in the observed differences.  Future
samples at both low and high redshifts should be able to answer this
question.


\section{Conclusions}\label{s:conc}

Using samples of high-quality low-redshift and intermediate-redshift
SN~Ia spectra, we have provided evidence of differences between the
SNe~Ia in these samples.  By using SN and host-galaxy photometry, we
are able to properly separate the SN and host-galaxy spectra.
Constructing a composite spectrum from 17 of our host-galaxy
contamination-corrected Keck/SDSS SNe, we see that the
intermediate-redshift SNe have $>20\%$ more UV flux ($\lambda <
3400$~\AA)of their low-redshift counterparts.  This UV excess is also
apparent in synthesized broad-band colors at high significance.  After
an exhaustive search for systematic errors, we determined that poor
spectral reductions, flawed post-processing techniques, or
Malmquist-biased sample selection are unlikely to be creating this
effect.  It is possible, however, that a combination of several small
systematic effects creates the UV excess.  Despite having otherwise
very similar properties, the UV flux of the Keck/SDSS sample does not
appear to be representative of the larger SDSS-II sample.

We have also shown that there may be a slight difference in the UV
spectra of SNe from low-mass and high-mass host galaxies (or,
alternatively, those that are actively forming stars versus those that
are not).  This difference is currently of low statistical
significance and requires larger samples to further improve confidence
in the result.  In particular, SNe with more massive hosts must be
probed at high redshift.  If SNe~Ia originating from different stellar
populations have similar optical light-curve shapes but different UV
spectra, observing the UV spectra of SNe~Ia may offer a new window to
understanding the progenitors of SNe~Ia and could potentially improve
distance estimates.  Several studies have now found a significant
correlation between host-galaxy mass, morphology, and star-formation
rate with Hubble residuals.  Detailed analysis of spectra may reveal
the physical cause of these correlations.

The UV ratio, $\mathcal{R}_{UV}$, has a substantially different
correlation with $M_{V}$ in the low-redshift and Keck/SDSS samples.
Since these samples are relatively small and that of Keck/SDSS has
little diversity in SNe~Ia, it is still unclear whether this
difference is inherent to low and high-redshift samples or is the
result of insufficient samples.  More observations at both low and
high redshifts are necessary to make this determination.

The differences in the UV spectra of low and high-redshift SNe~Ia may
be the result of SN~Ia evolution.  Future observations and careful
modeling are necessary to determine the magnitude and implications of
this possibility.  However, a more reasonable explanation is that the
low-redshift and high-redshift samples differ in some significant way.
We have accounted for phase and light-curve shape when comparing these
samples, but given the heterogeneous nature of (and incomplete
information for) the low-redshift sample, we were unable to account
for additional potential differences such as host-galaxy properties.
Since the low-redshift SNe~Ia were discovered through surveys which
target the most luminous galaxies in the nearby Universe and the
SDSS-II survey was untargeted, it is likely that the host-galaxy
properties for these samples are quite different.

The UV excess seen in the Keck/SDSS sample is not seen in the full
SDSS-II sample, as inferred from their light curve
\citepalias{Kessler09:cosmo}.  In fact, the full SDSS-II sample is, on
average, UV deficient relative to the low-redshift sample.  There is
no clear reason for this difference, and it may simply be the result
of a relatively small sample (although the Keck/SDSS sample represents
12\% of the full \citetalias{Kessler09:cosmo} SDSS-II sample).

A mismatch in the UV SEDs of SNe~Ia like that seen between the
Keck/SDSS and the low-redshift samples has important implications for
the estimated distances to all SNe~Ia.  Although the Keck/SDSS sample
is relatively large, it still is not fully representative of the
larger SDSS-II sample.  This indicates that there is a significant
scatter in the UV properties of SNe~Ia (at least at intermediate
redshift).  A simplistic treatment of this flux difference suggest
that the incorrect normalization of the peak luminosity (an offset in
$M_{V, {\rm peak}}$ for a given light-curve shape) and an incorrect
determination of the host-galaxy extinction (changing the measurement
of $A_{V}$) should both contribute to an underestimate of the correct
distance for an SN~Ia.  Fitting multi-band data that cover the
rest-frame UV and optical will tend to cause bluer SNe~Ia to have a
higher measured peak luminosity than reality, causing an opposite (and
dominant) effect.  This effect is constrained to the (near) UV of an
SN~Ia SED, and by ignoring this region, one should obtain an unbiased
(by this effect) distance estimate.

A light-curve fitter trained on a low-redshift sample (MLCS2k2) and
applied to high-redshift samples would produce significantly different
distance measurements if the rest-frame $U$ band were included or
excluded.  This was not seen with a light-curve fitter which was
trained on the full (low and high-redshift) sample (SALT2).  If the
cause of the $U$-band anomaly was a mismatch in the UV SEDs for the
low and high-redshift samples, SALT2 would have a model SED close to
both the low and high-redshift samples and differences would be
minimized.

For the James Webb Space Telescope (JWST), Euclid, and the future
potential Wide-Field InfraRed Survey Telescope (WFIRST) mission, which
will observe SNe in the near-infrared, there should be a relatively
small systematic color bias to a redshift of \about 2.5.  Since
current designs do not plan to observe SNe~Ia beyond that redshift,
differences seen in our sample should have a relatively small effect
on JWST, Euclid, or WFIRST measurements.

\begin{acknowledgments} 
R.J.F.\ is supported by a Clay Fellowship.

The SDSS-II SN team supplied targets, photometry, and light-curve fits
for this project; in particular, G.\ Miknaitis provided some of the
necessary data and discussions.  We thank J.~M.\ Silverman for help
with some of the observations, and R.\ Chornock for useful discussions
and for performing additional spectral reductions of several objects.
We are grateful to M.\ Blanton for discussions about the
\texttt{kcorrect} software and J.\ Bloom for insightful comments on an
early draft of this work.  We thank the referee, M.\ Stritzinger, for
helpful comments.  The spectra of intermediate-redshift SNe in this
study were obtained at the W. M. Keck Observatory, which is operated
as a scientific partnership among the California Institute of
Technology, the University of California, and the National Aeronautics
and Space Administration (NASA); the Observatory was made possible by
the generous financial support of the W.~M.\ Keck Foundation. The
spectra of low-redshift SNe in this study were obtained with the 3~m
Shane telescope at Lick Observatory, which is owned and operated by
the University of California. We thank the Keck and Lick staffs for
their assistance with the observations.

A.V.F. is grateful for the hospitality of the Aspen Center for
Physics, where this paper was finalized during the January 2012
program on ``The Physics of Astronomical Transients.''  This research
was made possible by NSF grants AST--0443378, AST--0507475,
AST--0607485, and AST-0908886, as well as by NSF CAREER award
AST-0847157 to S.W.J. at Rutgers University. Additional support was
provided by U.S. Department of Energy (DOE) grant DE-FG02-08ER41562
and by the TABASGO Foundation.  We also acknowledge funding from
NASA/{\it HST} grant GO--10182 from the Space Telescope Science
Institute, which is operated by the Association of Universities for
Research in Astronomy, Inc., under NASA contract NAS5-26555.

Funding for the SDSS and SDSS-II has been provided by the Alfred
P. Sloan Foundation, the Participating Institutions, NSF, DOE, NASA,
the Japanese Monbukagakusho, the Max Planck Society, and the Higher
Education Funding Council for England. The SDSS Web Site is
http://www.sdss.org/.

The SDSS is managed by the Astrophysical Research Consortium for the
Participating Institutions. The Participating Institutions are the
American Museum of Natural History, Astrophysical Institute Potsdam,
University of Basel, University of Cambridge, Case Western Reserve
University, University of Chicago, Drexel University, Fermilab, the
Institute for Advanced Study, the Japan Participation Group, Johns
Hopkins University, the Joint Institute for Nuclear Astrophysics, the
Kavli Institute for Particle Astrophysics and Cosmology, the Korean
Scientist Group, the Chinese Academy of Sciences (LAMOST), Los Alamos
National Laboratory, the Max-Planck-Institute for Astronomy (MPIA),
the Max-Planck-Institute for Astrophysics (MPA), New Mexico State
University, Ohio State University, University of Pittsburgh,
University of Portsmouth, Princeton University, the United States
Naval Observatory, and the University of Washington.

{\it Facilities:} 
\facility{Keck~I (LRIS)},
\facility{Magellan:Clay (MagE)},
\facility{Sloan Digital Sky Survey}

\end{acknowledgments}

\bibliographystyle{fapj}
\bibliography{../astro_refs}

\appendix

\section{Galaxy Subtraction}\label{a:galsub}

\subsection{Photometry Matching and Spectral Warping}

The first method to correct for galaxy contamination makes no
assumptions about the quality of our spectrophotometry and is similar
to the method presented by \citet{Ellis08}.  We perform extractions of
our SNe~Ia with a fixed aperture and background regions which avoid
any host-galaxy light.  Our resulting spectra are a combination of SN
and host galaxy with no attempt to remove host-galaxy contamination.

As a byproduct of the SMP photometry, we have the broad-band
photometry for the host galaxy at the position of the SN.  Using
multicolor light-curve shape \citep[MLCS2k2;][]{Jha07} template light
curves, we are able to interpolate the SMP photometry (independently
in each band) to determine the broad-band photometry for the SN at the
time the spectrum was obtained.  Using the photometry of both the SN
and the background at the position of the SN, we warp our spectrum to
match the measured photometry of the SN and background.  We do the
warping using both a spline and linear fit with three pivot points at
the flux-weighted centers of the \gri filters.  Since the SEDs are
slightly different for each object, we allow the pivot points to shift
in wavelength to match the true center of the filter for each
spectrum.  We can then subtract the reconstructed galaxy SED (properly
scaled to the galaxy photometry) from the spectrum, yielding a
spectrum which matches the photometry of the SN.

Since a spline will exactly fit the measured photometry, we perform a
Monte Carlo simulation to determine if the spline fitting is producing
reasonable residual spectra.  To do this, we randomly choose
photometry for each object assuming Gaussian distributions for the
photometry with a mean corresponding to the measured photometry and a
width corresponding to the uncertainty in the measurements.  Since the
uncertainties in the \gri\ bands are of similar magnitude, the average
of the residuals from the Monte Carlo simulations is typically very
similar to the spline using only the measured photometry.

We find that the photometry is always well fit by a linear function of
wavelength resulting in $\chi^{2}/\dof \ll 1$, indicating that for
many SNe this method is overfitting the data (i.e., no galaxy
subtraction is required for some SNe).  For most cases the slope of
the line is $\lesssim 1 \sigma$ (and in all cases the slope is $< 3
\sigma$) from 0, again indicating little to no correction required.
For our entire sample, there is no trend in the slope for a given
night, standard star, or slit position angle, with nearly equal
numbers of SNe with positive and negative slopes, suggesting that
there is no systematic problem affecting our galaxy subtraction.

\subsection{Color Matching}

The second method initially assumes that our spectrophotometry is well
calibrated (i.e., errors from the spectrophotometry do not dominate
the error in comparing the spectrum to the photometric colors).  The
flux calibration is verified both by the photometry matching above and
also after correcting for host-galaxy contamination.  For this method,
we attempt to remove as much galaxy contamination as possible when
extracting our spectra.  Although removing galaxy light at the point
of extraction prevents matching to the absolute scaling of photometry,
it tries to reduce the introduction of errors by subtracting a
potentially imprecise galaxy SED from a spectrum.  Since the galaxy
photometry determines the reconstructed galaxy SED, if our
spectrophotometry is well calibrated, then subtracting the
reconstructed galaxy SED from the observed spectrum such that the
colors of the residual spectrum matches those of the SN will result in
an SN-only spectrum.

We can demonstrate this mathematically as follows. In general, an
observed SN spectrum is defined by
\begin{equation}
  f_{\text{spec}} = A \left ( f_{\text{SN}} + Bf_{\text{gal}} \right ),
\end{equation}
where $f_{\text{spec}}$, $f_{\text{SN}}$, and $f_{\text{gal}}$ are the
vectors of fluxes in the observed spectrum, SN spectrum, and galaxy
spectrum, respectively, and $A$ and $B$ are normalization factors.
One can think of $A$ as normalizing the spectrum in an absolute sense
to account for slit losses, clouds, and other effects.  The parameter
$B$ can be thought of as a color normalization, where it is adjusted
until $f_{\text{spec}}$ and $f_{\text{SN}} + Bf_{\text{gal}}$ have the
same color.

As a byproduct of the SMP photometry, we have $p_{\text{gal}}$, the
broad-band photometry (in flux units) for the host galaxy at the
position of the SN.  Using multicolor light-curve shape
\citep[MLCS2k2;][]{Jha07} template light curves, we are able to
interpolate the SMP photometry (independently in each band) to
determine $p_{\text{SN}}$, the broad-band photometry (in flux units)
for the SN at the time of the spectrum.

We can define the function which translates spectra to synthetic
broad-band photometry as $P$, where $P(f_{\text{SN}}) = p_{\text{SN}}$
and $P(f_{\text{gal}}) = p_{\text{gal}}$.  Note that we impose the
first relationship, while the second relationship is required by our
method of determining $f_{\text{gal}}$.

From our spectrum, we are able to determine $P(f_{\text{spec}})$, the
broad-band synthetic photometry (in flux units) of the spectrum, which
includes both SN and galaxy light.  These vectors obey the equation
\begin{equation}\label{e:phot}
  P(f_{\text{spec}}) = A \left ( p_{\text{SN}} + Bp_{\text{gal}} \right ).
\end{equation}
\noindent
For this equation to be valid, we make two assumptions.  The first,
already noted above, is that our spectra are well flux calibrated.
The second assumption is that $B$, the relative fraction of the galaxy
and SN light in the observed spectrum, does not vary strongly with
wavelength.  We will return to both of these assumptions below.

Having \gri photometry for the galaxy and the SN, as well as the
observed spectrum and the reconstructed galaxy spectrum, we are able
to determine the factors $A$ and $B$.  We can then use these factors
to appropriately remove the reconstructed galaxy spectrum from the SN
spectrum,
\begin{equation}
  f_{\text{SN}} = A^{-1} f_{\text{spec}} - Bf_{\text{gal}}.
\end{equation}

\noindent
We have applied this technique to all SN spectra.  Figure~\ref{f:spec}
displays two of our observed spectra, the reconstructed host-galaxy
spectra, and the galaxy-contamination corrected SN spectra (see the
online version for figures showing all spectra).

\subsection{Galaxy Subtraction Performance}

We have applied two methods to properly correct for host-galaxy
contamination and to recover the SN SEDs for our sample of SNe~Ia.
One method attempts to match spectra to the measured brightnesses,
while the other strives to match to the measured colors.  Neither
technique is superior for all cases, but we will discuss their
performance for our data below.

The greatest advantage of the photometry-matching method is that it
can minimize errors from poor spectrophotometry.  By definition, the
spline-fitting version yields SN SEDs that exactly match the SN
photometry, while the line-fitting version will never increase the
residuals.  However, this method has two main drawbacks.  First, it
makes no attempt to reduce galaxy contamination during extraction.
Second, the warping of the spectrum is somewhat arbitrary and must be
extrapolated to wavelengths beyond those covered by our broad-band
photometry.

In a situation where a galaxy has a constant SED at all positions and
a smoothly varying surface-brightness profile, it is simple to remove
the galaxy contamination in a spectrum during extraction.  By
measuring the galaxy light on either side of the position of the SN in
the two-dimensional spectrum and interpolating over the position of
the SN, one can cleanly remove all galaxy contamination.  If, however,
one makes no attempt to remove the galaxy contamination during
extraction, but rather relies on a reconstruction of the galaxy SED
from broad-band photometry, one will almost certainly introduce
additional errors in the spectrum.

If our spectrophotometry is incorrect for a given object, the correct
function to warp the spectrum to match the photometry can never be
perfectly determined.  For our sample, we have assumed that this
function is either a spline or a line with anchors at the center of
the \gri bandpasses with fluxes equal to those measured from the \gri
photometry.  Although the corrections were typically small, deviations
from the fit are possible for wavelengths between the anchor points.
Additionally, any wavelengths beyond the anchor points must be
extrapolated.  Since our wavelength coverage (\about 3200--9200~\AA)
is much larger than the span from the bluest to the reddest anchor
points (approximately 4800 and 7500~\AA\ for $g$ and $i$,
respectively), we must either extrapolate for a significant portion of
our spectra or trim our spectra.  For the spline fitting, the implied
corrections in the extrapolation region can be large, while the
corrections from the linear fitting are more realistic.  For the
spline-fitting version, we have chosen to extrapolate to the blue edge
of the $g$ band and the red edge of the $i$ band, 3950~\AA\ and
8220~\AA, respectively.  For these wavelengths, we have some
information contributing to the photometry, while also limiting our
extrapolation.  For the linear-fitting version, we extrapolate
further, to 3200~\AA\ and 9000~\AA.

The one key advantage of the color-matching method is that galaxy
subtraction is performed during the extraction of the spectra.  This
will reduce any error associated with having an incorrect galaxy SED.
Again, this method requires two assumptions: the spectrophotometry is
not the dominant source of error and the PSF of our object does not
vary wildly with wavelength in a way where our parameter $B$ becomes
strongly dependent on wavelength.

We can examine our spectrophotometry by comparing synthetic broad-band
colors of our spectra to colors from the photometry.  In particular,
we are most interested in the relative spectrophotometry of our
spectra without any attempt to remove galaxy contamination during
spectral extraction.  These spectra can be directly compared to the
photometry of the SN and galaxy at the position of the SN.

\begin{figure}
\begin{center}
\includegraphics[angle=90,width=0.49\textwidth]{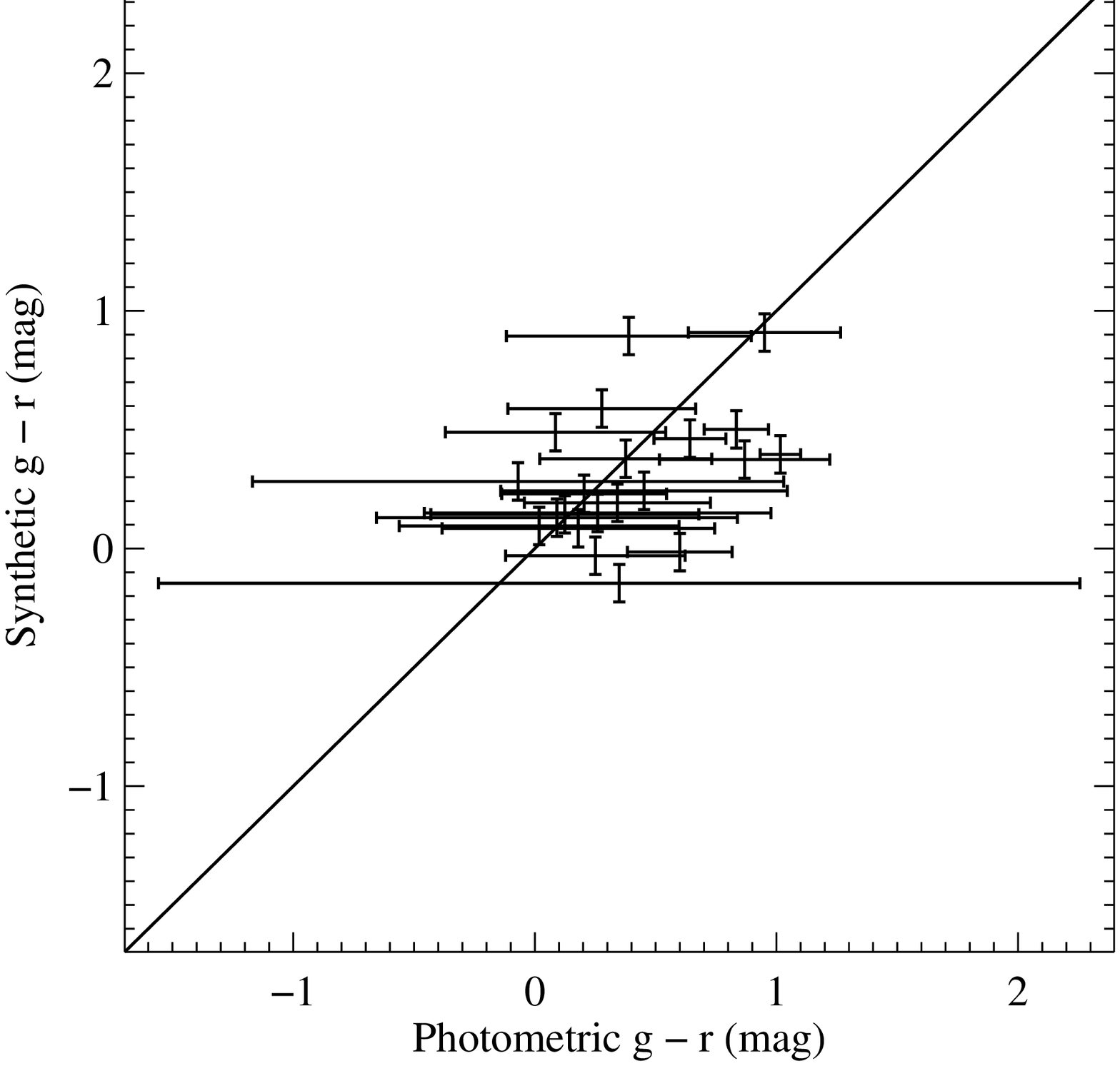}
\includegraphics[angle=90,width=0.49\textwidth]{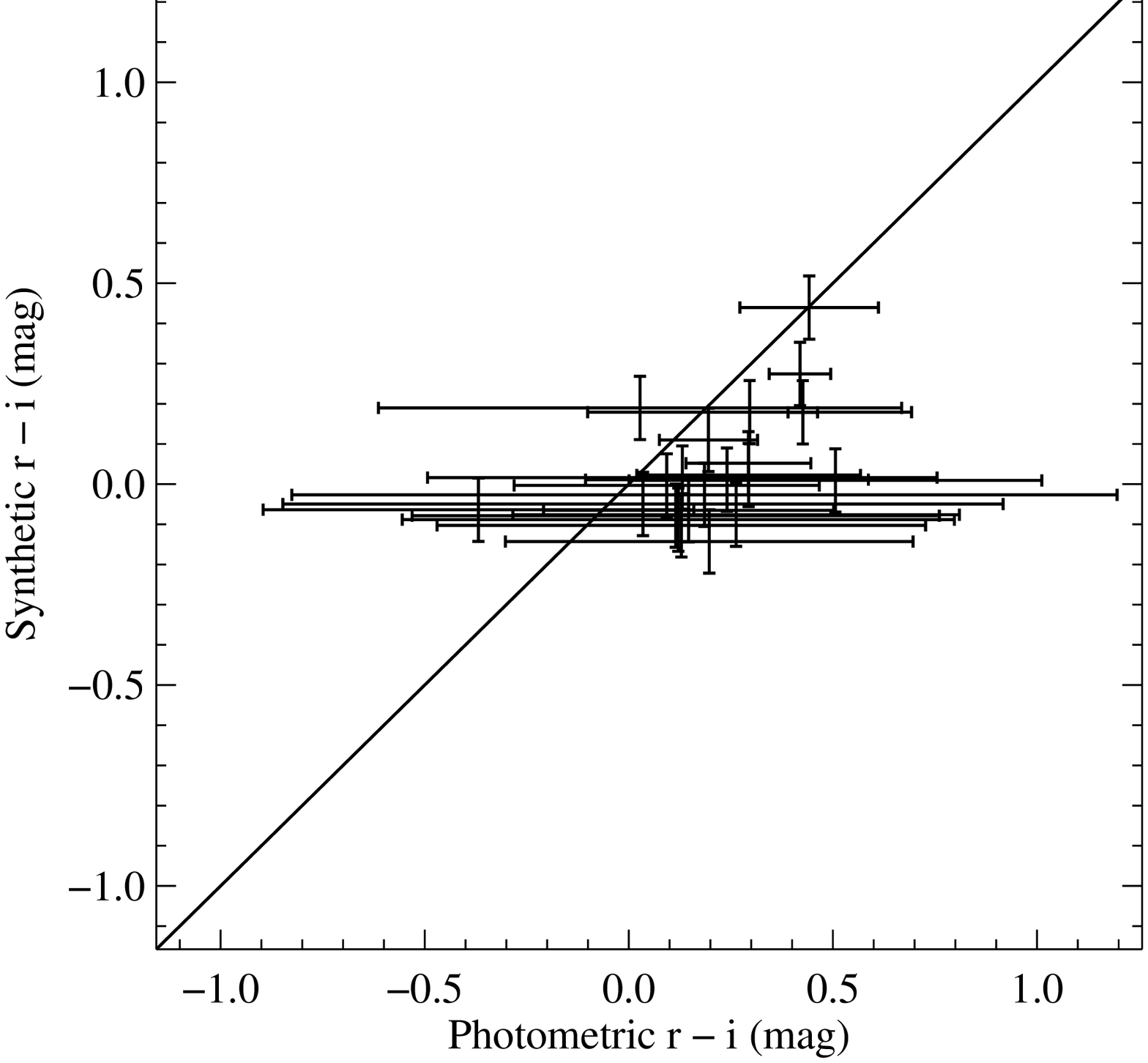}
\caption{({\it Left})  Synthetic $g-r$ colors of our SN spectra with
no attempt to remove galaxy contamination compared to the photometric
$g-r$ colors determined by adding the photometry of the SN at the time
of the spectrum to that of the galaxy at the position of the SN.
Overplotted is the line indicating equivalent photometric and
synthetic colors.  ({\it Right}) Synthetic $r-i$ colors of our SN
spectra with no attempt to remove galaxy contamination compared to the
photometric $r-i$ colors determined by adding the photometry of the SN
at the time of the spectrum to that of the galaxy at the position of
the SN.  Overplotted is the line indicating equivalent photometric and
synthetic colors.  The assumption that our spectrophotometry is
correct yields $\chi^{2}/\dof$ = 21.15/21.}\label{f:specphot}
\end{center}
\end{figure}

In Figure~\ref{f:specphot}, we present the $g-r$ and $r-i$ colors one
would find by measuring the fluxes in an aperture centered on the SNe
at the time of our spectra with no host-galaxy subtraction in our
images, as well as the synthetic colors of our spectra without any
attempt to remove galaxy contamination during spectral extraction.
These measurements are consistent with $\chi^{2}/\dof$ = 21.15/21,
showing that our spectrophotometry is very good over the bands where
we can measure it for these SNe.

\begin{figure}
\begin{center}
\epsscale{0.5}
\rotatebox{90}{
\plotone{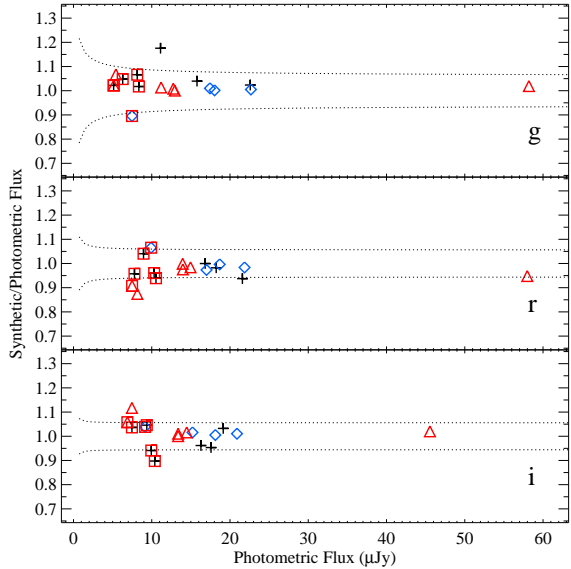}}
\caption{Ratio of synthetically created fluxes to measured photometric
fluxes for the Nominal sample.  The black crosses, blue diamonds, and
red triangles correspond to SNe with no, low ($< 8\%$), and high ($>
8\%$) host-galaxy contamination, respectively.  SNe where the galaxy
subtraction did not fail, but are at redshifts $0.27 < z < 0.33$, have
been marked with an additional red square.  The dotted lines represent
an estimate of the expected error from a combination of photometry,
spectral-flux calibration, and galaxy-subtraction
errors.}\label{f:galsub}
\end{center}
\end{figure}

A further test of our spectrophotometry is to examine the synthetic
photometry of our SN spectra after the galaxy contamination has been
removed.  In the case of the photometry-matching technique, the
residuals are by design small.  For the color-matching method, on the
other hand, the residuals are not guaranteed to be small.  If our
spectrophotometry (or alternatively, our galaxy SEDs) were incorrect,
we could obtain a minimal residual in the colors that are still far
from the measured values for the SN.  In Figure~\ref{f:galsub}, we
present the ratio of the synthetic fluxes recovered after galaxy
subtraction to the measured photometric fluxes for each SN in our
sample.  Ignoring SNe~Ia with low S/N or failed galaxy-template
determination, the average of the residuals is 1.03, 0.97, 1.01, with
a scatter in the residuals of 5.4\%, 4.5\%, and 5.0\% in the $g$, $r$,
and $i$ bands, respectively.  If we also remove the SNe with $0.27 < z
< 0.33$, the average of the residuals is 1.03, 0.97, 1.01, with a
scatter in the residuals of 5.1\%, 3.7\%, and 4.2\% in the $g$, $r$,
and $i$ bands, respectively.  These uncertainties are consistent with
the errors expected from the spectrophotometry, photometry, and galaxy
subtraction (\citealt{Matheson08}; Silverman et~al., submitted;
\citealt{Blanton03}).  For the photometry-matched, linearly corrected
spectra including (excluding) SNe with $0.27 < z < 0.33$, the average
of the residuals is 1.06, 1.06, and 1.08 (1.04, 1.03, and 1.07) with
scatter in the residuals of 6.0\%, 9.0\%, and 11.8\% (3.0\%, 6.6\%,
and 9.4\%) for the $g$, $r$, and $i$ bands, respectively.

From Figure~\ref{f:galsub}, we can also infer that our parameter $B$
does not vary significantly with wavelength.  If, for instance, $B$
were a monotonically increasing function of wavelength, but we were
not accounting for it, we would expect to oversubtract our galaxy SED
in the $g$ band and undersubtract our galaxy SED in the $r$ band.
Since the average values of the ratio of the synthetic to photometric
flux are consistent with 1.0 in all bands, we conclude that $B$ is not
strongly dependent on wavelength.

From our tests, we find that both the photometry-matching and
color-matching methods produce excellent galaxy-contamination
corrected SN spectra over the \gri bands.  Because of the
extrapolation of the spline function, we do not believe our spline-fit
photometry-matched spectra can be extended blueward of the $g$ band or
redward of the $i$ band; similarly, the linear-fit photometry-matched
spectra are limited to 3200--9000~\AA.  Considering the excellent
spectrophotometry of our spectra for the \gri bands and our experience
with $U$-band flux calibration for many low-redshift SNe~Ia (Silverman
et~al., submitted), the color-matching technique should yield properly
flux-calibrated spectra redward of \about 3200~\AA.  We have performed
our analysis using both the photometry-matched and color-matched
spectra, yielding similar results.  We will present some results from
both methods, but will focus on the color-matched spectra.

\subsection{Galaxy Subtraction Failures}

An SN fails the galaxy subtraction when the best-fit galaxy SED has
larger flux values than the galaxy-contaminated SN spectrum for a
significant portion of the spectrum.  This can occur if there is
incorrect galaxy or SN photometry or if the SN spectrum is not
properly flux calibrated.  Additionally, for SNe with $0.27 < z <
0.33$, where incorrect galaxy SEDs are possible, the reconstructed SED
may be unphysical.

There are four SNe~Ia for which the galaxy subtraction fails.  Three
of these (SNe~16644, 16789, and 20829) have redshifts of \about 0.3
and significant galaxy contamination.  The unphysical SEDs of
reconstructed galaxy spectra at $z \approx 0.3$ are the obvious reason
for a failure of the fit.  Despite this, by constraining a heavily
smoothed version of the residual SN spectrum to have non-negative
fluxes at all places besides the ends and at emission lines (which may
be oversubtracted), we can achieve a rough, but probably
oversubtracted, SED of the SN.  Although this is not a rigorous
treatment of the galaxy contamination, it could be useful for
visualization.

SN~7512 also fails the galaxy-subtraction procedure.  It has
relatively low S/N spectra, making galaxy subtraction less precise.
The four SNe with failed galaxy subtraction have been removed from our
sample for any analysis, leaving a sample of 17 SNe~Ia that we use for
our analysis.  Additional spectroscopy of the SN hosts could improve
the SEDs of the SNe with $0.27 < z < 0.33$, but the faintness of the
unobserved host galaxies makes this task difficult with current
facilities.

\section{Comparison to SNLS $z \approx 0.5$ Composite Spectrum}

Two recent studies generated composite spectra of high-redshift
SNe~Ia.  \citet{Foley08:comp} presented composite spectra generated
from high-redshift SNe~Ia from the ESSENCE survey \citep{Miknaitis07}.
Unfortunately, the lack of multi-color information prevented
galaxy-contamination correction similar to that implemented in this
study.  As a result, there was little information on changes to the SN
SED.  However, a second study by \citet{Ellis08} of SNLS SNe~Ia did
present composite spectra after correcting for the galaxy
contamination.  The two spectra (one ``early'' spectrum with $t <
-4$~days and one ``maximum'' spectrum with $-4 < t < 4$~days; both
with $\mean{z} \approx 0.5$) can provide an interesting comparison to
our composite spectra.

\begin{figure}
\begin{center}
\epsscale{0.5}
\rotatebox{90}{
\plotone{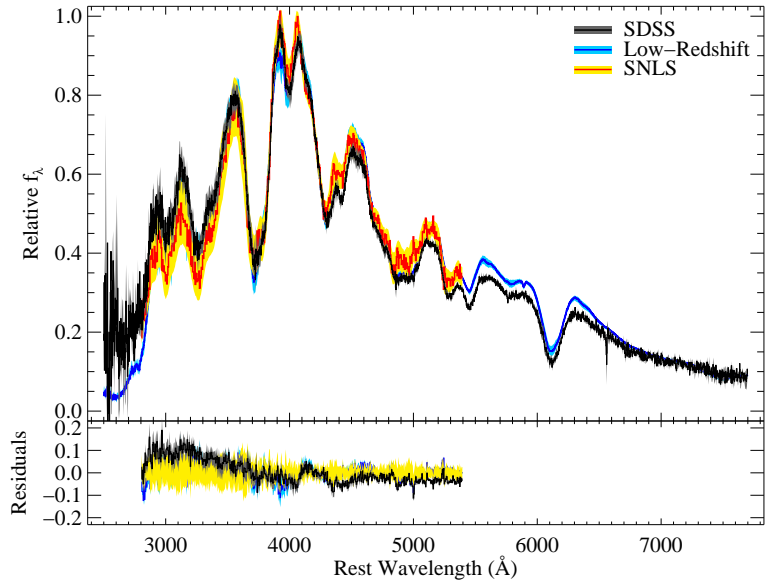}}
\caption{({\it Top panel}):  Composite spectrum created from our
Keck/SDSS sample (black curve) compared to the SNLS maximum-light
composite spectrum from \citet{Ellis08} (red curve), and the
maximum-light low-redshift composite spectrum (blue curve).  The grey,
yellow, and light-blue regions are the 1$\sigma$ boot-strap sampling
errors for the Keck/SDSS, SNLS, and low-redshift composite spectra,
respectively.  ({\it Bottom panel}): The yellow region is the
1$\sigma$ boot-strap sampling region for the SNLS composite spectrum.
The black and blue curves are the residuals of the SNLS and the
Keck/SDSS (Keck/SDSS minus SNLS) and low-redshift (low redshift minus
SNLS) composite spectra, respectively.  The grey and light-blue
regions are the residuals of the SNLS composite spectrum and the
Keck/SDSS and low-redshift 1$\sigma$ boot-strap sampling regions,
respectively.}\label{f:ellis}
\end{center}
\end{figure}

In Figure~\ref{f:ellis}, we illustrate the Keck/SDSS, SNLS, and
low-redshift maximum-light composite spectra.  For this comparison, we
scaled the spectra to the SNLS spectrum over its entire wavelength
range.  The SNLS spectrum is consistent with both the Keck/SDSS and
low-redshift spectra over the wavelength range sampled.  The early and
maximum-light SNLS spectra both have a median stretch of 1.06, which
corresponds to $\Delta = -0.23$.  Therefore, the SNLS SNe have broader
light curves (and are correspondingly more luminous SNe~Ia) than those
of the Keck/SDSS or low-redshift samples.  The UV SEDs of SNe~Ia are
strongly dependent on their light-curve shape (see
Section~\ref{ss:delta}), and since we do not know the average
light-curve shape of the composite spectrum (the individual spectra
have different weights and therefore contribute different amounts to
the composite spectrum), the comparison is of less utility than that
of the low-redshift and Keck/SDSS composite spectra.  Despite these
differences, it is worth noting that the SNLS spectrum has less flux
than the Keck/SDSS spectrum (but still within the error region) and
slightly more flux than the low-redshift spectrum at most wavelengths
(again, still within the error region).  Larger differences may exist
in these samples, but since the SNLS spectrum does not probe much of
the optical region (stopping blueward of 5400~\AA), the lever arm of
scaling the spectra to match at long wavelengths is unavailable for
this analysis.  As seen with the low-redshift and Keck/SDSS spectra in
Figure~\ref{f:ellis}, matching the spectra at only these blue
wavelengths can cause large divergence at redder wavelengths.

\begin{figure}
\begin{center}
\epsscale{0.5}
\rotatebox{90}{
\plotone{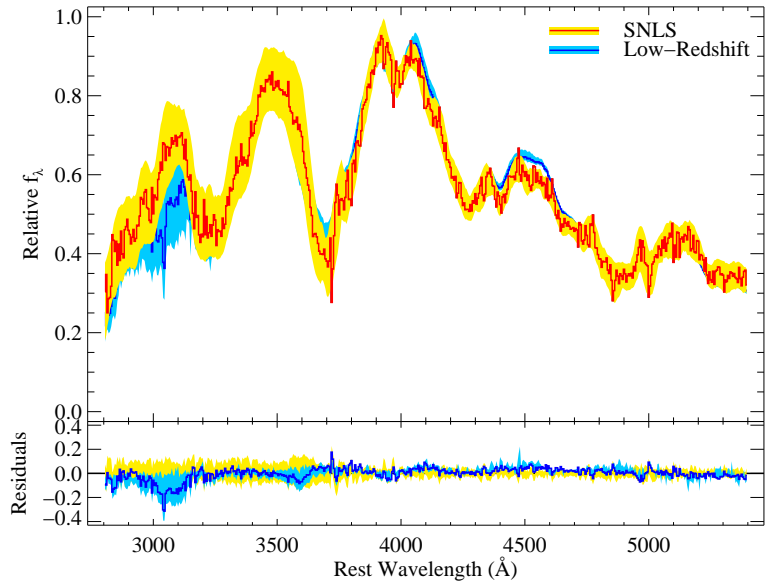}}
\caption{({\it Top panel}):  SNLS ``Early'' composite spectrum created
from \citet{Ellis08} (red curve; $t < -4$~days) compared to the
pre-maximum-light low-redshift composite spectrum from
\citet{Foley08:comp} (blue curve, $\mean{t} = -6.6$~days).  The yellow
and light-blue regions are the 1$\sigma$ boot-strap sampling errors
for the SNLS and low-redshift composite spectra, respectively.  ({\it
Bottom panel}): The yellow region is the 1$\sigma$ boot-strap sampling
region for the SNLS composite spectrum.  The blue curve is the
residual of the SNLS and the low-redshift composite spectra.  The
light-blue region is the residual of the SNLS composite spectrum and
the low-redshift 1$\sigma$ boot-strap sampling
regions.}\label{f:early_ellis}
\end{center}
\end{figure}

In Figure~\ref{f:early_ellis}, we present a comparison of the SNLS
``Early'' composite spectrum with a low-redshift $\mean{t} =
-6.6$~days composite spectrum from \citet{Foley08:comp}.  Over the
restricted wavelength range, the two spectra appear to be consistent.
There is a large deviation at \about 3100~\AA, with the SNLS spectrum
having excess flux compared to the low-redshift spectrum, but the
uncertainties in both the SNLS and low-redshift spectra are large
enough for the spectra to be consistent.  Similar to the maximum-light
spectra, these samples likely have different light-curve shape
characteristics, and additional information is necessary to make
definitive conclusions.

The expectation is that any physical change in SN~Ia spectra with
redshift would be monotonic with redshift.  With that in mind, the
fact that the SNLS composite spectrum is slightly above the
low-redshift composite spectrum and slightly below the Keck/SDSS
composite spectrum is puzzling.  Perhaps the Keck/SDSS sample is
biased in a way we have not been able to detect, or our assumption of
monotonic spectral behavior is incorrect.  However, it is worth noting
that when splitting the Keck/SDSS sample by redshift, the
higher-redshift composite spectrum has less UV flux than the
lower-redshift composite spectrum.  Although this difference is not
significant given our boot-strap sampling errors, it shows the same
trend as with the SNLS spectra.

Without combining the individual SNLS SN~Ia spectra in the same manner
as the Keck/SDSS SN~Ia spectra, it is difficult to quantify the
similarities and differences.  In particular, \citet{Ellis08} warped
the spectra to match their photometry before correcting for
host-galaxy contamination.  The reasoning for this is that
differential slit losses and atmospheric extinction cause errors in
the spectrophotometry.  We have shown that our reduction methods
reproduce $B-V$ colors with a low dispersion ($\sigma = 0.063$~mag)
for low-redshift SNe~Ia \citep{Matheson08}, and we see that there is a
low scatter in the synthesized photometry (Figure~\ref{f:galsub}) of
our sample of objects with no discernible host galaxy in the images
(despite there possibly being some underlying host contributing
slightly to the spectrum).  Hence, we conclude that our
spectrophotometry is reasonable and has uncertainties similar to those
associated with host-galaxy contamination.  Furthermore, the spectral
warping of \citet{Ellis08} extrapolates into the rest-frame UV for the
lower-redshift ($z \lesssim 0.3$) objects.


\end{document}